\documentclass[a4paper,11pt]{article}
\usepackage[dvips]{graphicx}
\usepackage{amssymb}
\usepackage{amsmath}
\usepackage{cite}

\begin{document}

\title{Late Time Cosmological Evolution in f(R) theories with Ordinary and Collisional Matter}
\author{
V.K. Oikonomou$^{1,2}$\,\thanks{voiko@physics.auth.gr}\\
Department of Theoretical Physics, Aristotle University of Thessaloniki,\\
54124 Thessaloniki, Greece$^1$\\
Max Planck Institute for Mathematics in the Sciences\\
Inselstrasse 22, 04103 Leipzig, Germany$^{2}$
\\\\
N. Karagiannakis\,\thanks{nikar@auth.gr}\\
Polytechnic School, Aristotle University of Thessaloniki,\\
54124 Thessaloniki, Greece
} \maketitle

\begin{abstract}
We study the late time cosmological evolution of $f(R)$ theories of modified gravity, with the matter content of the universe being that of collisional self interacting matter. We assume that the universe is described by a flat Friedmann-Lemaitre-Robertson-Walker metric and that it is matter and dark energy dominated. The results of our numerical analysis for a collisional matter $f(R)$ theory are compared with those resulting from pressure-less matter $f(R)$ theory and from the $\mathrm{\Lambda}\mathrm{CDM}$ model. As we shall demonstrate, the resulting picture can vary from model to model, indicating that the effect of collisional matter in $f(R)$ theories is strongly model dependent. Particularly, in a few cases, may give better fit to the $\mathrm{\Lambda}\mathrm{CDM}$ model. In all studied cases, the effective equation of state parameter does not cross the phantom divide, both in the collisional matter and pressure-less matter $f(R)$ theories. Finally, we thoroughly study the effects of collisional matter on one of the $f(R)$ models that is known to provide a unified description of early time inflation and late time acceleration. The overall picture of the evolution of the universe is not drastically affected, apart from the matter era which is further enhanced with an additional matter energy density term, which is of leading order. However, a fully consistent description of the universe's evolution requires the introduction of a dark energy compensate in the total energy density, a concept very well known from the literature.
\end{abstract}

\section*{Introduction}

Recent results dated back in the beginning of the 21st century have altered the way the universe evolution 
was perceived before, in a drastic way, owing to the discovered late time acceleration that currently
the universe undergoes \cite{riess}, a result verified by observing supernovae Ia. This result in
addition to the very recent (2013) verification of the inflationary era of the early universe
\cite{bicep}, have brought into play many alternative theories of gravity in order to consistently
describe, in a unified and consistent theoretical framework, early time and late time acceleration of the universe.
Up to now, the observational data seem to favor a flat universe filled with pressure-less matter
(dust formed matter) and also with a non-zero cosmological constant. Particularly, according to the
new Planck telescope observational data for the present epoch, the universe is consistently
described by the $\mathrm{\Lambda}\mathrm{CDM}$ model, and consists of ordinary matter
($\Omega_m\sim4.9\%$), cold dark matter ($\Omega_{DM}\sim
26.8\%$) and also dark energy ($\sim 68.3\%$). The late time acceleration is mostly attributed to 
the latter, that is, dark energy and one of the current research objectives is to model this negative
pressure fluid consistently.

The concept of dark energy can be described in an elegant and self-consistent way by the $f(R)$ 
modified theories of gravity and related modifications. For an important stream of reviews and
important articles on this vast research topic, the reader is referred to
\cite{reviews1,reviews101,reviews2,reviews3,reviews4,reviews5,reviews6,reviews7,reviews8,reviews9,
importantpapers1,importantpapers2,importantpapers3,importantpapers4,importantpapers5,
importantpapers6,importantpapers7,importantpapers8,importantpapers9,importantpapers10,importantpapers11,
importantpapers12,importantpapers13,importantpapers14,importantpapers15,importantpapers16,importantpapers17,
importantpapers18,importantpapers19,importantpapers20,importantpapers21,importantpapers22,
importantpapers23,importantpapers24,importantpapers25,importantpapers26} and references therein. In
addition, some of these $f(R)$ theory models can describe simultaneously late time acceleration and
inflation, see for example \cite{sergeinojirimodel} , where this was explicitly done by Nojiri and Odintsov. 
Apart from $f(R)$ theories of gravity, there exist alternative theories that can also describe dark energy, but in a different context, see for example \cite{capo,capo1,peebles,faraonquin,tsujiintjd,tsagas} and the book of Tsujikawa \cite{reviews10}.
In $f(R)$ theories of modified gravity, what actually changes in comparison to the Einstein
equations of General Relativity is not the left hand side of the Einstein equations, but the right
hand side, this why sometimes the dark energy is called geometric. In order to have late time
acceleration, it is required that the theory contains a negative $w$ fluid, which can in some way be
consistently incorporated in the energy momentum tensor of the theory. This feature naturally
appears in all viable $f(R)$ theories and therefore late time acceleration solutions of the
Friedmann-Robertson-Walker equations, occur in these modified gravity theories
\cite{reviews1,reviews101,reviews2,reviews3,reviews4,reviews5,reviews6,reviews7,reviews8,reviews9,   
importantpapers1,importantpapers2, importantpapers3,importantpapers4,importantpapers5,
importantpapers6,importantpapers7,importantpapers8,importantpapers9,importantpapers10,importantpapers11,
importantpapers12,importantpapers13,importantpapers14,importantpapers15,importantpapers16,importantpapers17,
importantpapers18,importantpapers19,importantpapers20,importantpapers21,importantpapers22,
importantpapers23,importantpapers24,importantpapers25,importantpapers26,sergeinojirimodel}. However,
any modified gravity theory has to be confronted with the successes of General Relativity, and
therefore there exist many constraints that have to be satisfied in order a model can be considered
viable, at least to some extend. Particularly, quite stringent constraints are imposed to $f(R)$ theories from
planetary, star formation and local tests of General Relativity (see for example
\cite{reviews1,reviews101,reviews2,importantpapers4,importantpapers7}). In addition, in most cases what is
actually the primary objective for the $f(R)$ models is to have concordance with the $\mathrm{\Lambda}\mathrm{CDM}$ model and quite significant works exist in the literature towards this direction
\cite{importantpapers3,importantpapers7,importantpapers10,importantpapers22}, see also the work of Basilakos and Perivolaropoulos \cite{basilakos1,basilakos2}. Moreover, every $f(R)$ theory has a scalar-tensor Jordan frame 
counterpart theory with $\omega$ zero and non-zero potential, and this theory in turn has an Einstein
frame counterpart theory. In order an $f(R)$ theory is viable, the Einstein frame scalaron has to be
classical, so that quantum mechanical stability of the theory is ensured (see for example
\cite{reviews1,reviews101,reviews2,reviews3,reviews4,reviews5} and related to the subject references therein).
Solutions in various strong curved gravitational backgrounds were given in
\cite{solutions1,solutions2,solutions3,solutions4} and also modified gravitational theories with
non-minimal curvature-matter coupling, were studied thoroughly in
\cite{bertolami01,bertolami0,bertolami1,bertolami2,bertolami3,bertolami4} and references therein. 

When late time acceleration was discovered, one of the largest mysteries was to explain the 
coincidence problem, that is, why the universe started to accelerate at such a relatively low value
of the cosmological redshift parameter $z$. Particularly, the exact value of $z$, at which the
transition from deceleration to acceleration occurs is called transition redshift and we will denote
it ''$z_t$''. The $f(R)$ dark energy models can predict this transition from deceleration to
acceleration in a successful way and also there exist studies to describe the passage from a matter
domination epoch to late time acceleration, see \cite{importantpapers3}. Also for a informative
study of universe's evolution in $f(R)$ theory, see the work of Appleby and Battye \cite{importantpapers21,importantpapers22} and also the work of Mukherjee and Banerjee \cite{importantpapers26}. In
view of the coincidence problem and the transition from deceleration to acceleration, the purpose of
this paper is to study how the late time evolution of the universe is affected in a matter and dark energy
epoch, with matter being collisional and hence self-interacting in some way. Particularly, the focus is on 
how the whole process is affected, that is, if the universe accelerates more slowly or faster, and
also how the transition redshift is affected with the addition of collisional self-interacting matter.
Note that we are interested in the post-recombination era, at which the universe is filled with
matter and dark energy (late time evolution of the universe). The models of matter we shall use, appear already in the literature
\cite{kleidis,freese,gondolo,card} and shall briefly describe them in the following sections. The
transition from deceleration to acceleration is mostly affected by the $f(R)$  theory and hence the
inclusion of this self-interacting matter can affect, as we will see, the rate of the process and the
exact value of the transition redshift, but not the qualitative behavior of the evolutionary
process. The final result however is strongly model dependent. We believe that this option of having an alternative form of matter, instead of cold dark matter for example, can free up the way we choose the unspecified variables of the various $f(R)$
models, owing to the fact that the model parameters can have less stringent values. This is quite
useful, since in many cases in the literature, and at the expense of concordance with the experimental
value of the transition redshift, the model parameters were chosen in such a way that the model may
loose its viability (see for example the tachyonic instability of the exponential model in reference
\cite{importantpapers26}). As we demonstrate shortly in the following sections, we will be able to
offer more freedom to the choice of model parameters. Let us note finally that, our approach does not
imply any interaction between dark energy and dark matter, in terms of some generalized equation of
state, as it was done for example in \cite{elizaldegener1,elizaldegener11,elizaldegener12,elizaldegener2,elizaldegener3,elizaldegener4,
elizaldegener6} and references therein. In our case, there is no interaction between matter and dark
energy and the only existent interactions are the self interactions of matter, expressed in terms
of a logarithmic or a power-law dependence on the matter-energy density. Since in most of the sections, what is actually examined is the late time acceleration era of the universe, but it's worthwhile to investigate all the evolution eras that the universe will experience, from inflation to late time acceleration. This is done in the last section of the article, in which we study in detail the effect of collisional matter to all evolution eras of the universe, within the context of $f(R)$ theories. In addition, we make use of a very well known technique introduced for the first time in \cite{importantpapers3}, called dark energy compensate, in order to provide a fully correct evolution for the universe. Our version of the dark energy compensate, takes into account all the contributions of the collisional matter. 

This paper is organized as follows: In section 1, we briefly present the essentials of $f(R)$ 
theories, in section 2 we present the collisional matter model we will use following closely
the work of Kleidis and Spyrou \cite{kleidis}, and also present the cosmological
evolution in the context of two $f(R)$ theories in a universe filled with collisional matter. The models we used in section 2, where thoroughly studied in \cite{importantpapers26} by Mukherjee and Banerjee, so we can directly compare the collisional matter $f(R)$ theories with the results of reference \cite{importantpapers26}. In section 3, we repeat the analysis corresponding to collisional matter,
by using a form of matter dictated by the so-called Cardassian cosmological models, always in the
context of $f(R)$ theories. In section 4, we examine very well known and viable $f(R)$ models, and 
we explicitly demonstrate how the deceleration parameter and the effective equation of state is
affected by the existence of collisional matter. In section 5 we provide a detailed analysis on how the collisional matter affects the universe's evolution during the inflation, matter domination and late time acceleration eras, by using a very well known $f(R)$ model, which describes the three eras successfully. In addition, we use the dark energy compensate to provide a fully correct description of the universe's evolution. The conclusions along with a discussion on the results and perspective
future work, follow in the end of the paper.

\section{Essentials of $f(R)$ Theories in the Jordan Frame}

In order to render the article self-contained, we review in brief the basic
features of $f(R)$ gravity theories in the Jordan frame using the theoretical framework of the metric
formalism. For an important stream of review papers and articles on the subject see
\cite{reviews1,reviews101,reviews2,reviews3,reviews4,reviews5,reviews6,reviews7,importantpapers1,
importantpapers2,importantpapers3,importantpapers4,importantpapers5,importantpapers6,
importantpapers7,importantpapers8,importantpapers9,importantpapers10,importantpapers11,importantpapers12,importantpapers13} and
references therein. 

We assume a pseudo-Riemannian geometrical background on the manifolds, which locally is a Lorentz metric (which is the FRW metric in our case). Moreover, we consider a torsion-less, symmetric, and
metric compatible affine connection, the so-called Levi-Civita connection. Working on such geometric
backgrounds, the Christoffel symbols are equal to:
\begin{equation}\label{christofell}
\Gamma_{\mu \nu }^k=\frac{1}{2}g^{k\lambda }(\partial_{\mu }g_{\lambda \nu}+\partial_{\nu
}g_{\lambda \mu}-\partial_{\lambda }g_{\mu \nu})
\end{equation} 
and moreover the Ricci scalar becomes:
\begin{equation}\label{ricciscalar}
R=g^{\mu \nu }(\partial_{\lambda }\Gamma_{\mu \nu }^{\lambda}-\partial_{\nu }\Gamma_{\mu \rho
}^{\rho}-\Gamma_{\sigma \nu }^{\sigma}\Gamma_{\mu \lambda }^{\sigma}+\Gamma_{\mu \rho }^{\rho}g^{\mu
\nu}\Gamma_{\mu \nu }^{\sigma}).
\end{equation}
The $f(R)$ theories of modified gravity provide a generalization of the standard the Einstein-Hilbert 
action, with the four dimensional action of $f(R)$ theories in the Jordan frame being equal to:
\begin{equation}\label{action}
\mathcal{S}=\frac{1}{2\kappa^2}\int \mathrm{d}^4x\sqrt{-g}f(R)+S_m(g_{\mu \nu},\Psi_m),
\end{equation}
In the above relation (\ref{action}) $\kappa$ is $\kappa^2=8\pi G$ and also $S_m$ stands for the matter action containing the matter fields $\Psi_m$. For the sake of simplicity, in this section it shall be assumed that the form of the $f(R)$ theory that will be used
is $f(R)=R+F(R)$ and as already mentioned, that the metric formalism framework is used. 

By varying action (\ref{action}) with respect to the metric tensor $g_{\mu \nu}$, we obtain the following equations of motion:
\begin{equation}\label{eqnmotion}
f'(R)R_{\mu \nu}(g)-\frac{1}{2}f(R)g_{\mu \nu}-\nabla_{\mu}\nabla_{\nu}f'(R)+g_{\mu \nu}\square
f'(R)=\kappa^2T_{\mu \nu}.
\end{equation} 
In the above, the prime of the $f(R)$ function denotes differentiation with respect to the Ricci scalar, that is $f'(R)=\partial f(R)/\partial R$ and in addition $T_{\mu \nu}$ stands for the energy 
momentum tensor. 

One of the most striking features of the $f(R)$ modified gravity theories is that, what essentially renders them modified with regards to the Einstein-Hilbert theory of gravity is the direct modification of the right hand side of the Einstein equations, with the left remaining unaltered. The equations of motion
(\ref{eqnmotion}) for $f(R)$ theories can be cast in the following form:
\begin{align}\label{modifiedeinsteineqns}
R_{\mu \nu}-\frac{1}{2}Rg_{\mu \nu}=\frac{\kappa^2}{f'(R)}\Big{(}T_{\mu
\nu}+\frac{1}{\kappa}\Big{[}\frac{f(R)-Rf'(R)}{2}g_{\mu \nu}+\nabla_{\mu}\nabla_{\nu}f'(R)-g_{\mu
\nu}\square f'(R)\Big{]}\Big{)}.
\end{align}
Thereby, we obtain an additional contribution for the energy momentum tensor, originating from the term:
\begin{equation}\label{newenrgymom}
T^{eff}_{\mu \nu}=\frac{1}{\kappa}\Big{[}\frac{f(R)-Rf'(R)}{2}g_{\mu
\nu}+\nabla_{\mu}\nabla_{\nu}f'(R)-g_{\mu \nu}\square f'(R)\Big{]}.
\end{equation}
This term (\ref{newenrgymom}) models actually the dark energy in $f(R)$ theories of modified gravity, and hence in these theories dark energy has a geometric origin. By taking 
the trace of equation (\ref{eqnmotion}), we obtain straightforwardly the following equation:
\begin{equation}\label{traceeqn}
3\square f'(R)+R f'(R)-2f(R)=\kappa^2 T,
\end{equation}
with $T$ being the trace of the energy momentum tensor $T=g^{\mu \nu}T_{\mu \nu}=-\rho+3P$,
and, in addition $\rho_m$ and $P_m$ stand for the total matter-energy density and pressure respectively. 

By observing equation (\ref{traceeqn}) we can see that in $f(R)$ theories of gravity there exists another degree of freedom, described by the function $f'(R)$. This degree of freedom is commonly known as the scalaron field and the equation of motion of this field is equation (\ref{traceeqn}). Finally, in this paper we shall assume a flat Friedmann-Lemaitre-Robertson-Walker spacetime of the form,
\begin{equation}\label{metricformfrwhjkh}
\mathrm{d}s^2=\mathrm{d}t^2-a^2(t)\sum_i\mathrm{d}x_i^2
\end{equation}
with the Ricci scalar in this background being equal to:
\begin{equation}\label{ricciscal}
R=-6(2H^2+\dot{H}),
\end{equation}
In the above equation, $H(t)$ stands for the Hubble parameter and the ``dot'' indicates differentiation with respect to time. We have to note that for the purposes of this article we shall consider the Hubble parameter as a function of $z$ and also that the Ricci scalar and all relevant quantities shall be expressed as a function of the redshift parameter $z$.

\section{A Logarithmic Model of Self-interacting Matter and Late Time Cosmological Evolution in $f(R)$ 
Theories}

\subsection{Brief Review of the Logarithmic Model for Collisional Matter}

In this section we briefly introduce the model of collisional matter that we shall use. In
order to make the argument more clear, we denote with ''$\varepsilon_m$'' the total
mass-energy density corresponding to matter, instead of using the common notation $\rho_m$. The
reason for this shall be clear soon. Notice that, this $\varepsilon_m$ is the $T^{00}$ component
of the energy momentum tensor. If we assume that the matter is not collisional (pressure-less), then
$\varepsilon_m=\rho_m$, but in contrast, when matter is collisional, the energy momentum tensor
receives another contribution in terms of potential energy which we denote as $\Pi$, that includes
all the extra interactions between the collisional matter, expansions and compressions mainly (we
follow the notation of \cite{kleidis} and of the book of Fock \cite{fock}, pages 90-93). Adding the
potential energy $\Pi$, the total energy density $\varepsilon_m$ is then given by the following
expression \cite{kleidis,fock}:
\begin{equation}\label{generenergydensit}
 \varepsilon_m=\rho_m+\rho_m\Pi
\end{equation}   
We have to note that $\rho_m$ refers to that part of the energy momentum tensor which does not 
change due to the hydrodynamic flows (or in the course of motion) of the gravitational fluid that
describes the matter \cite{kleidis,fock}. In addition, the term $\rho_m\Pi$ actually expresses the
energy density part of the energy momentum associated with thermodynamical content of the
collisional matter. This gravitational fluid is obviously not dust, but has a positive pressure and
satisfies the following equation of state:
\begin{equation}\label{eqnstate}
 p_m=w\rho_m
\end{equation} 
with $\rho_m$ as we clearly stated earlier, the rest mass density, that is, the part of the total 
energy $\varepsilon_m$ that remains unaffected by the internal motions of the gravitational cosmic
fluid. The parameter $w$ takes values between zero and one, that is:
\begin{equation}\label{eqnstatewval}
0\leq w \leq 1
\end{equation}
with $w=0$, the non-collisional matter case, in which case $\varepsilon_m=\rho_m$. So actually the 
values of $w$ dictate that the universe actually consists of a mixture of ordinary non-collisional
and collisional mater, when $0<w<1$. Notice also that when $w=1$, the matter content is that of a
scalar field without a scalar potential. The motions of the volume elements in the interior of a
continuous medium are governed by the continuity equation:
\begin{equation}\label{continuityeqn}
 T^{\mu \nu}_{;\nu}=0
\end{equation}
with the energy momentum in our case taking the following form:
\begin{equation}\label{continuityeqnenergymom}
 T^{\mu \nu}=(\varepsilon_m+p_m)u^{\mu}u^{\nu}-p_mg^{\mu \nu}
\end{equation}
where $u^{\mu}$ is the four velocity at some position of the fluid's volume element, $g^{\mu \nu}$ 
the metric tensor of the universe and also $\varepsilon_m$ the total energy density of the
gravitational fluid. The form of the potential energy density expressing collisional matter is of
the following form \cite{kleidis}:
\begin{equation}\label{potenerg}
\Pi=\Pi_0+w\ln (\frac{\rho_m}{\rho_0})
\end{equation}
with $\rho_0$ and $\Pi_0$ denote present values of the motion invariant mass energy density and of 
the potential energy, respectively. Therefore, the total energy density of the gravitational fluid
is equal to:
\begin{equation}\label{totenergydens}
\varepsilon_m=\rho_m\Big{[}1+\Pi_0+w \ln (\frac{\rho_m}{\rho_0})\Big{]}
\end{equation}
Notice that the total mass-energy density has the usual $\rho_m$ term at the beginning plus a 
constant potential term $\Pi_0$ plus the logarithmic interaction term. Hence, the interaction terms
are the ones containing the potential term $\Pi_0$ plus the logarithm term. Moreover, owing to the
continuity equation for the gravitational fluid (\ref{continuityeqn}), the continuity equation of
the gravitational fluid in a spatially flat FRW metric is:
\begin{equation}\label{contquenne}
\dot{\varepsilon}_m+3\frac{\dot{a}}{a}(\varepsilon_m+p_m)=0
\end{equation}
with $a$ denoting the scale factor, which leads to \cite{kleidis}:
\begin{equation}\label{evlotuniomatt}
\rho_m=\rho_0\big{(}\frac{a_0}{a}\big{)}^3
\end{equation}
with $a_0$ the present value of the scale factor. The collisional matter is actually described by equations (\ref{totenergydens}) and 
(\ref{evlotuniomatt}) and we shall use these in the rest of this section. Note that the value of
$\Pi_0$ is equal to \cite{kleidis}:
\begin{equation}\label{pio}
\Pi_0=\Big{(}\frac{1}{\Omega_M}-1\Big{)}
\end{equation}
Let us note that a universe containing only collisional matter cannot explain the present 
acceleration of the universe, at least when someone assumes that the physics is described by the
Jordan frame. In reference \cite{kleidis}, the authors gave a different interpretation of the
evolution by using conformal transformations. For informative accounts on conformal transformations and related issues see the book of Faraoni \cite{faraoni} and also \cite{fujii}. We do not adopt their argument
of choosing a more physical frame at which matter is pressure-less, but we rather stay in the Jordan
frame (without discussing the up to date still debatable issues of choosing the most appropriate
physical frame), embed the collisional mater in an $f(R)$ framework and explicitly check the late time
evolution of the universe in this context.

\subsection{Late Time Cosmological Evolution in terms of the Deceleration Parameter}

Having in mind that one of the most challenging issues of the modern cosmology is to find a correct
physical explanation for the dynamical transition from deceleration to acceleration at late times, we study the late time evolution of the universe in the context of $f(R)$ theories with collisional matter
by examining the deceleration parameter $q(z)$. The latter has developed to be a quite useful tool
for late time cosmology, since it can be used as a consistency check of cosmological parameters
\cite{bedran} and in some cases it has been proposed to be a cosmological number itself \cite{lima}.
In reference to the last issue, the transition redshift has been pointed out that it can be a quite
useful cosmological probe \cite{lima}. In view of these facts, in the next two sections we shall
briefly give the expressions with regards to the deceleration parameter in $f(R)$ theories
containing non-zero pressure matter, and also for the $\mathrm{\Lambda}\mathrm{CDM}$ case.

\subsection{Deceleration Parameter in $f(R)$ Theories}

In order to derive an expression for the deceleration parameter, we shall follow the approach and 
notation of reference \cite{importantpapers26}, with the difference that in our case, matter has
pressure and also we have a generalized matter-energy density. For a quite detailed and informative article on these issues, see the work of Capozziello et al. \cite{importantpapers15} and related references therein. This will slightly modify the resulting
equations, compared to the ones obtained by Mukherjee and Banerjee in \cite{importantpapers26}. So we consider a general $f(R)$ theory
described by the action (\ref{action}). In a flat FRW background, the Einstein equations read (we
expand and re-write the ones quoted in the previous section for notational clarity and convenience):
\begin{align}\label{neweineqns}
&
3\frac{\dot{a}^2}{a^2}=\frac{\varepsilon_m}{f'}+\frac{1}{f'}\Big{[}\frac{1}{2}(f-Rf')-3\dot{R}
f''\frac{\dot{a}}{a}\Big{]} \\ \notag &
2\frac{\ddot{a}}{a}+3\frac{\dot{a}^2}{a^2}=\frac{p_m}{f'}-\frac{1}{f'}\Big{[}\ddot{R}f''+\dot{R}
^2f'''+2\dot{R}f''\frac{\dot{a}}{a}-\frac{1}{2}(f-Rf')\Big{]}
\end{align}
where as usual, dot and prime denote differentiation with respect to time and R, respectively. Also,
$\varepsilon_m$ and $p_m$ are the expressions for total mass energy density and matter pressure 
given in relations (\ref{eqnstate}) and (\ref{totenergydens}). Following \cite{importantpapers26} we
define $\rho_c$ and $p_c$ as follows:
\begin{align}\label{neweineqnscurvat}
& \rho_c=  \Big{[}\frac{1}{2}(f-Rf')-3\dot{R}f''\frac{\dot{a}}{a}\Big{]} \\ \notag &
p_c= \frac{1}{f'}\Big{[}\ddot{R}f''+\dot{R}^2f'''+2\dot{R}f''\frac{\dot{a}}{a}-
\frac{1}{2}(f-Rf')\Big{]}
\end{align}
which are the contribution of the curvature (dark energy) to the total energy density and pressure respectively. 
Moreover, using the contracted Bianchi identity along with relations (\ref{neweineqnscurvat}), we get:
\begin{equation}\label{contrbianch}
\frac{\mathrm{d}}{\mathrm{d}t}\Big{(}\frac{\varepsilon_m+\rho_c}{f'}\Big{)}+3H\Big{(}\frac{
\varepsilon_m+p_m+\rho_c+p_c}{f'}\Big{)}=0
\end{equation}
But having in mind that the total matter-energy density, due to the continuity equation of the 
energy momentum tensor, satisfies the generalized equation (\ref{contquenne}), the above equation
(\ref{contrbianch}), can be simplified and brought into the following form:
\begin{equation}\label{simplifeqns}
18\frac{f''}{f'}H(\ddot{H}+4H\dot{H})+3(\dot{H}+H^2)+\frac{f}{2f'}+\frac{\varepsilon_m}{f'}=0
\end{equation}
The above equation (\ref{simplifeqns}), can be written in terms of the redshift z, and by using 
equation (\ref{totenergydens}), we end up to the following equation:
\begin{align}\label{basiceqntiontobeused}
& \frac{\mathrm{d}^2H}{\mathrm{d}z^2}=\frac{3}{(1+z)}\frac{\mathrm{d}H}{\mathrm{d}z}-
\frac{1}{H}\Big{(}\frac{\mathrm{d}H}{\mathrm{d}z}\Big{)}^2  \\ \notag &
-\frac{3f'\big{(}H^2-(1+z)H\frac{\mathrm{d}H}{\mathrm{d}z}\big{)}+\frac{f}{2}+\rho_{m0}(1+z)^3\Big{(
}1+\Pi_0+3w \ln (1+z)\Big{)}}{18(1+z)^2H^3f''}
\end{align}
Equation (\ref{basiceqntiontobeused}) shall be our main tool for analyzing the late time cosmological 
evolution of various viable $f(R)$ models we shall study in the following sections. We shall perform a numerical
analysis in order to find the Hubble parameter $H(z)$ and then, by using the expression:
\begin{equation}\label{expression1}
q(z)=\frac{(1+z)}{H(z)}\frac{\mathrm{d}H}{\mathrm{d}z}-1
\end{equation}
we shall study the dependence of the deceleration parameter as a function of the redshift. We are
mainly interested in the transition from deceleration to acceleration and also on when this happens,
that is, finding the transition redshift $z_t$. Moreover, we shall investigate how this transition
redshift changes when the model parameters change. Moreover, we also study the dependence of the
effective equation of state parameter $w_{eff}$, defined as:
\begin{equation}\label{effeqnofst}
w_{eff}=\frac{p_c}{\rho_c+\varepsilon_m}
\end{equation}
as a function of the redshift. We compare the results we shall find with the results coming from 
pure $f(R)$ theories with pressure-less matter and also with the results coming from the
$\mathrm{\Lambda}\mathrm{CDM}$ model. It worths giving the expression for the deceleration
parameter corresponding to the $\mathrm{\Lambda}\mathrm{CDM}$  model, for a flat universe, which is
\cite{bedran}:
\begin{equation}\label{decelLCDM}
q(z)=\Big{(}\Omega_{m0}\frac{(1+z)^3}{2}-L_0\Big{)}\Big{(}\Omega_{m0}(1+z)^3+L_0\Big{)}
\end{equation}
with $\Omega_{m0}=0.279\pm 0.015$ and $L_0=0.721\pm 0.015$ \cite{bedran}. The initial conditions 
and the values of the parameters we use in the numerical study are $\Pi_0=2.58423$, $q(0)=-0.81$, $\frac{\mathrm{d}H}{\mathrm{d}z}=0.19$, using the conventions of references \cite{importantpapers26,kleidis}.

\subsection{Study of two General $f(R)$ Models with Collisional Matter}

\begin{figure}[h]
\centering
\includegraphics[width=15pc]{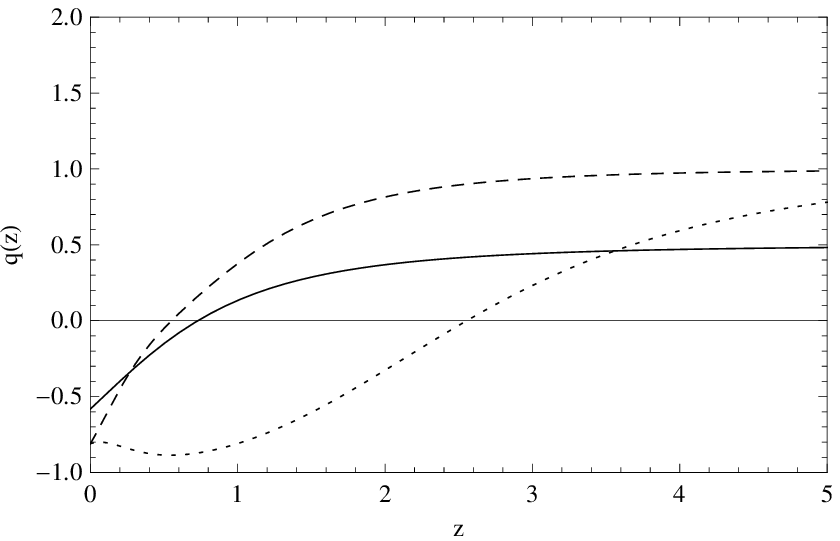}
\includegraphics[width=15pc]{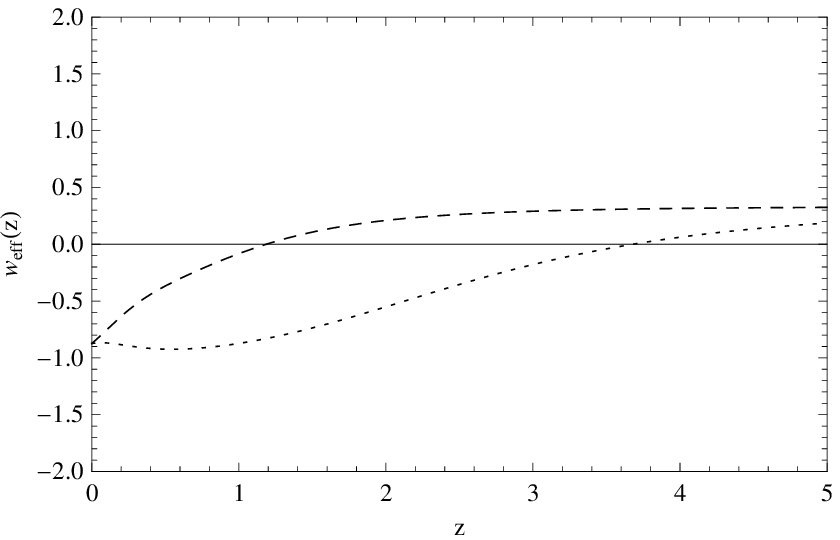}
\caption{Plots of $q(z)$ and $w_{eff}$ over z for the model
$f(R)=\lambda_0(\lambda+R)^{n}$. The dashed, dotted, solid lines refer to collision-less, collisional, and
$\Lambda\mathrm{CDM}$ models, respectively.}\label{fig:banerjee1plots}
\end{figure}

In order to make contact with the existing literature and have a direct comparison of the different physics that the (logarithmic) collisional self-interacting matter brings along, we will study the two models proposed in \cite{importantpapers26}. The first one is a modified power-law model the actual form of which is:
\begin{equation}\label{eq:banerjee1}
  f(R)=\lambda_0(\lambda+R)^n,
\end{equation}
with $\lambda_0$, $\lambda$, and $n$ being positive constants. We choose the following values of the parameters, namely, $\lambda_0=1$, $\lambda=13.5$, and finally $n=0.5$. We have performed a numerical analysis of the cosmological equations of motion (\ref{basiceqntiontobeused}) for the $f(R)$ model (\ref{eq:banerjee1}) with ordinary pressure-less matter ($w=0$) and collisional matter ($w=0.6$) and we now discuss in detail the results.  In Fig. \ref{fig:banerjee1plots} we have presented the functional dependence of the deceleration parameter $q(z)$ and of the effective equation of state $w_{eff}$ as a function of the redshift parameter $z$, for the $\Lambda\mathrm{CDM}$, the $f(R)$ model with and without collisional matter.  In both plots, the dashed, dotted, solid lines refer to collision-less, collisional, and $\Lambda\mathrm{CDM}$ models, respectively. As a first comment, we observe that the behavior of the $f(R)$ model without collisional matter is similar to the one obtained by the authors of (\ref{eq:banerjee1}). By looking Fig. \ref{fig:banerjee1plots}, we can see that the $f(R)$ model with collisional matter, actually behaves much more worse in reference to the $f(R)$ model with pressure-less matter, since the latter is much more close to the $\Lambda\mathrm{CDM}$ curve. In addition, the transition redshift $z_t$ corresponding to the $f(R)$ model with collisional matter is equal to approximately $z_\simeq 2.5$, which is a rather large value in reference to the one observed experimentally and of course this can be seen in (\ref{fig:banerjee1plots}). The same applies for the effective equation of state $w_{eff}$. The results obtained here hold true for a wide range of values for the $w$ parameter corresponding to the collisional matter. So in conclusion, the collisional matter in this case rather makes the final picture worse and has nothing interesting to offer in this particular model. Nevertheless, as we will see in the following sections the latter feature is strongly model dependent, since for some models, the effect of collisional matter is to make the final picture better. As a final comment, observe that $w_{eff}$ in Fig. \ref{fig:banerjee1plots}, ends to a value of $w_{eff}$, with $w_{eff}>-1$, so there is no crossing of the phantom divide \cite{phantom}.

\begin{figure}[h]
\centering
\includegraphics[width=15pc]{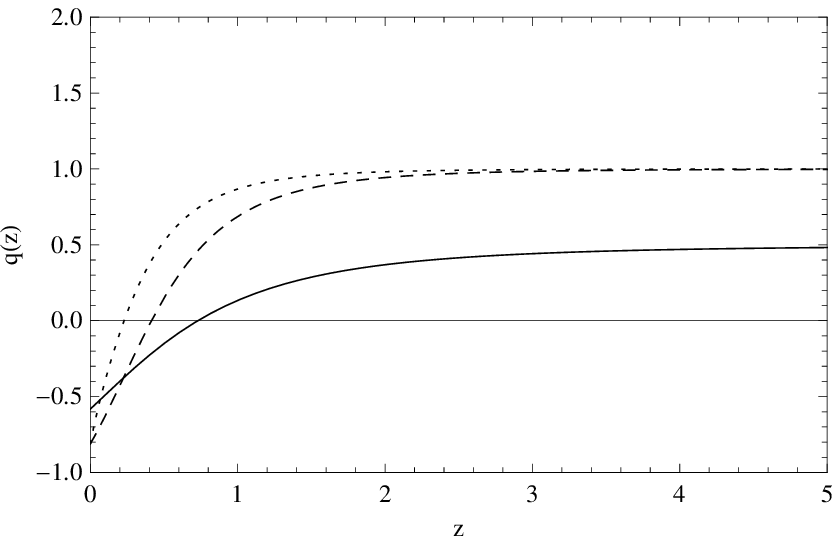}
\includegraphics[width=15pc]{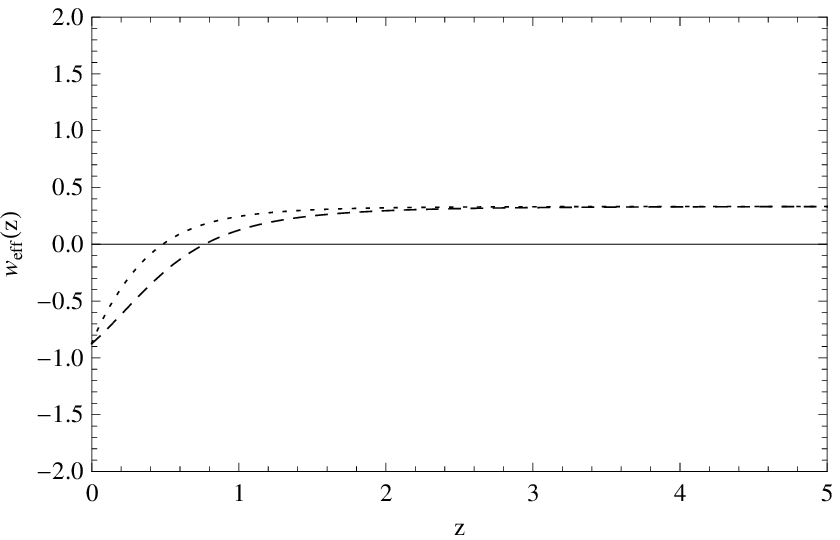}
\caption{Plots of $q(z)$ and $w_{eff}$ over z for the model
$f(R)=R_0e^{aR}$. The dashed, dotted, solid lines refer to collision-less, collisional, and
$\Lambda\mathrm{CDM}$ models, respectively.}\label{fig:banerjee2plots}
\end{figure}

The second model studied in reference \cite{importantpapers26} was an exponential $f(R)$ model, of the following form,
\begin{equation}\label{eq:banerjee2}
  f(R)=R_0e^{aR},
\end{equation}
where $R_0$ and $a$ are constant parameters. We chose their numerical values to be $R_0=1$ and $a=1.5$,
and we preformed a numerical analysis, the results of which are presented in Fig. \ref{fig:banerjee2plots}. By observing the behavior of $q(z)$ for both the collisional and collision-less case, we can see that in this case, collisional and collision-less curves are actually pretty close to each other, and both away from the $\Lambda\mathrm{CDM}$ curve. But the transition redshift for both cases is very close to the $\Lambda\mathrm{CDM}$ one. Exactly the same applies for the $w_{eff}$ behavior. In conclusion, in both $f(R)$ models studied in this section, there is the transition from deceleration to acceleration for both collision-less and collisional matter cases. In addition, the collisional matter $f(R)$ resulting curves may be similar or different form the non-collisional ones, a fact that is strongly model dependent.  

Before closing this section, we shall discuss some features and conventions related to the figures of this section and also applied for all the figures appearing in this paper. Mainly we discuss why we chose the redshift to vary from $z=0$ to $z=5$. In most cases in standard cosmology, the cosmological distances are determined by using standard candles, such as TypeIa supernovae or Gamma Ray Bursts (GRB hereafter). The TypeIa supernovae have a redshift that varies between $0<z<1.7$ \cite{ariadna}. On the other hand GRBs are visible in much more higher redshifts that stretches up to $z=6$ \cite{ariadna} and this is why we used $z=5$ to our plots, in order to take into account the future observational data coming form GRBs coming from high redshifts.

\section{Study of two General $f(R)$ Models with Cardassian Self-interacting Matter}

\subsection{Essentials of the Cardassian Model}

Apart from the collisional matter, there exist other proposals in the literature that also adopt the self-interacting matter approach \cite{freese,gondolo,card}. In this section we briefly review the related models and apply the results to the two models of reference \cite{importantpapers26} which we studied in the previous section. The assumption of another matter-energy density function $\rho $ was firstly adopted in \cite{freese,gondolo}. In such a scenario, matter has self interactions which are characterized by 
negative pressure and in the original model no vacuum energy was present. According to the so-called
Cardassian model of matter \cite{freese,gondolo,card}, the total energy density $\varepsilon_m$ of
matter is equal to:
\begin{equation}\label{card1}
\varepsilon_m=\rho+\rho_K(\rho)
\end{equation} 
where $\rho$ stands for ordinary matter-energy density and $\rho_K (\rho)$ is the term describing 
the new interacting matter and is, in general, a function of the ordinary mass-energy density
$\rho$. In the original Cardassian model, the form of this function $\rho_K(\rho)$ was of the
following form:
\begin{equation}\label{card2}
\rho_K(\rho)=B\rho^n
\end{equation}
and the equation of state which determines the relation between pressure and matter density was 
obtained by assuming the direct modification of the first Friedmann equation and also assuming
continuity in terms of the energy-momentum tensor. Moreover, the value of ''$n$'' was assumed to be
$n<2/3$, in order to have acceleration. 
\begin{figure}[h]
\centering
\includegraphics[width=15pc]{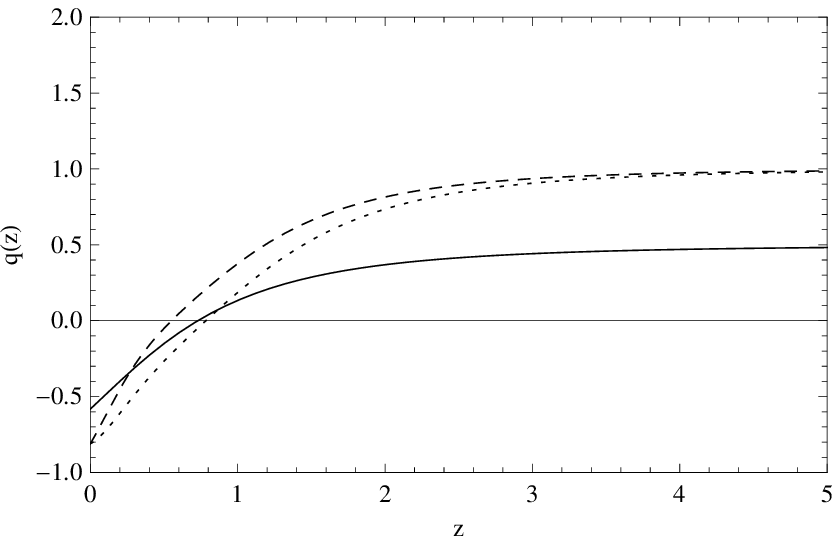}
\includegraphics[width=15pc]{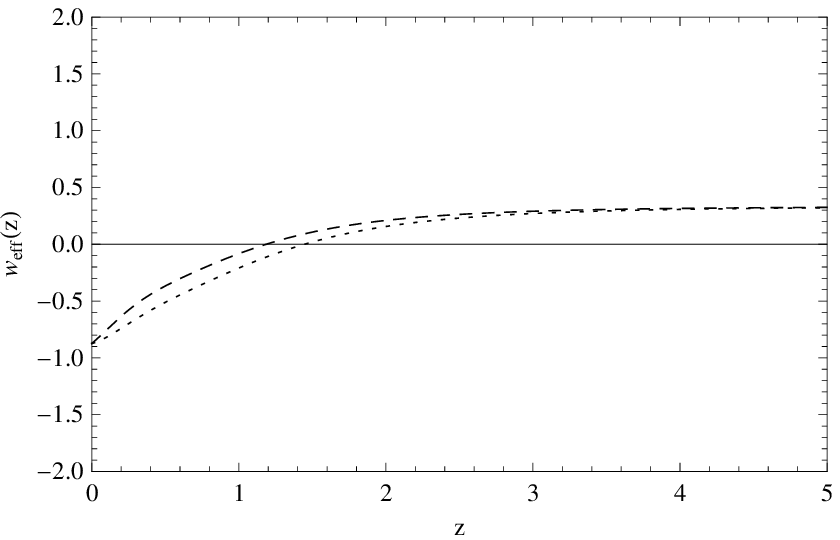}
\caption{Plots of $q(z)$ and $w_{eff}$ over z for the model
$f(R)=\lambda_0(\lambda+R)^{n}$. The dashed, dotted, solid lines refer to collision-less, Cardassian, and
$\Lambda\mathrm{CDM}$ models, respectively.}\label{fig:banerjee3plots}
\end{figure}
In this article, we assume however that the total
matter-energy density is described by relations (\ref{card1}) and (\ref{card2}), but we do not adopt
the technique used by the authors of \cite{freese,gondolo}, in order to determine the pressure, but
instead we assume that, as in the case studied in the preceding sections, the late time
evolution is driven by the geometric dark fluid with negative pressure originating from the $f(R)$
theories plus a gravitating fluid of positive pressure, satisfying the following equation of state:
\begin{equation}\label{card3}
p=w_k\rho
\end{equation}
with $\rho$ the rest mast energy density that is robust to self-interactions of matter. In addition,
 the total mass-energy density is given by the combination of relations (\ref{card1}) and
(\ref{card2}), that is:
\begin{equation}\label{card4}
\varepsilon_m=\rho+B\rho^n
\end{equation}
with $B$ being one of the free parameters of the theory. Moreover, we shall raise the constraint, 
$n<2/3$ allowing $n$ to be a free parameter of the theory, since late time acceleration can be
accomplished by the dark energy coming from the $f(R)$ theory. 
\begin{figure}[h]
\centering
\includegraphics[width=15pc]{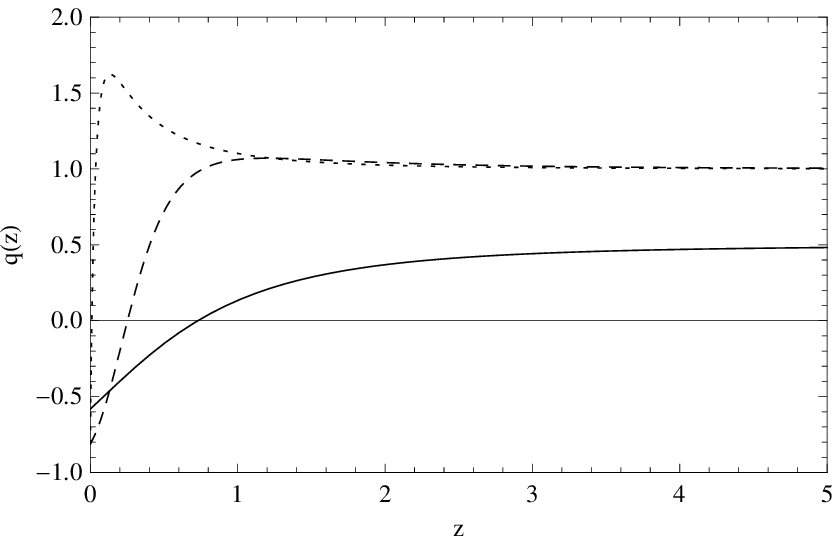}
\includegraphics[width=15pc]{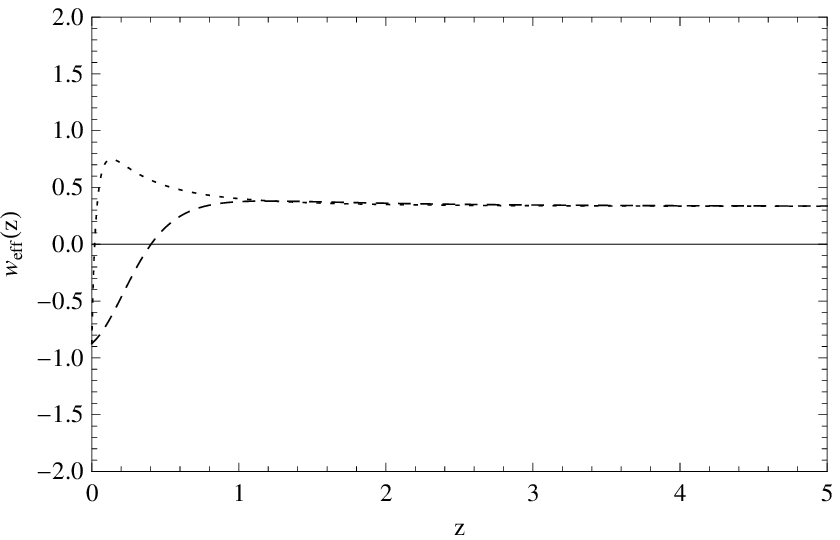}
\caption{Plots of $q(z)$ and $w_{eff}$ over z for the model
$f(R)=R_0e^{aR}$. The dashed, dotted, solid lines refer to collision-less, collisional, and
$\Lambda\mathrm{CDM}$ models, respectively.}\label{fig:banerjee4plots}
\end{figure}
In the rest of this section, we make
this assumption for the total mass density and incorporate the theory into an $f(R)$ theory framework and we thoroughly
examine the late time evolution of the universe in a matter and dark energy dominating universe. We use the
approach we adopted in the previous sections and compare the results with the
$\mathrm{\Lambda}\mathrm{CDM}$  model ones and also with the $f(R)$ model ones but with pressure-less
matter. The equation that gives the Hubble parameter as a function of the redshift in this case is
equal to:
\begin{align}\label{basiceqntiontobeusedcardassian}
& \frac{\mathrm{d}^2H}{\mathrm{d}z^2}=\frac{3}{(1+z)}\frac{\mathrm{d}H}{\mathrm{d}z}-
\frac{1}{H}\Big{(}\frac{\mathrm{d}H}{\mathrm{d}z}\Big{)}^2  \\ \notag &
-\frac{3f'\big{(}H^2-(1+z)H\frac{\mathrm{d}H}{\mathrm{d}z}\big{)}+\frac{f}{2}
+\rho_{m0}(1+z)^3\Big{(}1+B (\rho_{m0})^{n-1}(1+z)^{3(n-1)}\Big{)}}{18(1+z)^2H^3f''}
\end{align}
In the following, we numerically solve this equation for the models we used in the previous section 
and compare the results with the ones coming from the $\mathrm{\Lambda}\mathrm{CDM}$ along with
the ones coming from $f(R)$ theories with ordinary pressure-less matter. 

\subsection{Late Time Evolution of the Universe in two $f(R)$ Models with Cardassian Matter Content}

To exemplify the behavior of Cardassian matter, we test the $f(R)$ models appearing in relations
(\ref{eq:banerjee1}) and (\ref{eq:banerjee2}) and the results are given in Figs.
\ref{fig:banerjee3plots} and \ref{fig:banerjee4plots} respectively. For the first model we use the same set of parameter values as we used in the previous section and we now discuss the numerical results presented in Fig.
\ref{fig:banerjee3plots}. In addition we use $B=0.2$, $n=-3$. As a first comment, by looking the $q(z)$ plot, we observe that the Cardassian matter $f(R)$ theory has similar behavior to the one corresponding to pressure-less matter $f(R)$ theory. In addition, the Cardassian matter $f(R)$ theory is much more close to the $\mathrm{\Lambda}\mathrm{CDM}$ curve and more importantly, it's transition redshift is much more closer to the $\mathrm{\Lambda}\mathrm{CDM}$ one, a feature that the pressure-less matter $f(R)$ theory does not share. The same reasoning applies to the $w_{eff}$ plot. In reference to the second model appearing in relation (\ref{eq:banerjee2}), using the value $a=0.1$ and looking at the $q(z)$ plot of Fig. \ref{fig:banerjee4plots}, we observe that the Cardassian matter $f(R)$ theory is very far away from the $\mathrm{\Lambda}\mathrm{CDM}$ curve and also gives a transition redshift which is much more smaller than the $\mathrm{\Lambda}\mathrm{CDM}$ value. The same reasoning applies for the $w_{eff}$ too. Therefore, in the first model (\ref{eq:banerjee1}) the Cardassian matter $f(R)$ theory gives a better fit to the $\mathrm{\Lambda}\mathrm{CDM}$ curve in comparison to the pressure-less matter curve $f(R)$ theory. The converse occurs in second model (\ref{eq:banerjee2}). It is important to note that, as before, these $f(R)$ models have $w_{eff}$ which at small redshifts is negative but becomes positive for larger $z$. In addition, we have $w_{eff}>-1$, which is away from the phantom regime \cite{phantom}. Also, with respect to the exponential model (\ref{eq:banerjee2}), the transition occurs at $z>0.5$ in case of collision-less
matter $f(R)$ theory and much earlier for the Cardassian matter $f(R)$ theory, while
in the polynomial model (\ref{eq:banerjee1}), both cases give almost the same transition point.

\section{Study of Other Known $f(R)$ Models Late Time Cosmological Evolution with and without Self-Interacting Matter}\label{viability}

In this section we study the late time cosmological evolution of very well known viable $f(R)$ cosmological models \cite{reviews1,reviews101,reviews2,reviews3,reviews4,reviews5,reviews6,reviews7,reviews8,reviews9} with the addition of self-interacting matter to these. In addition, we shall compare the behavior of $q(z)$ and $w_{eff}$ of these models, with $f(R)$ models containing pressure-less matter and with the $\mathrm{\Lambda}\mathrm{CDM}$ model. We critically test whether the addition of self-interacting matter of collisional type has anything new to offer in the $f(R)$ theories theoretical framework. 

Before starting the study, it worths recalling when an $f(R)$ theory is considered to be viable. As we mentioned in the introduction, there exist very stringent constraints coming from local gravity tests, cosmological observations and quantum mechanical stability \cite{reviews1,reviews101,reviews2,reviews3,reviews4,reviews5,reviews6,reviews7,reviews8,reviews9}. The most important of these constraints  are actually the following:
\begin{itemize}
  \item $f'(R)>0$ for $R\geq R_0>0$, where $R_0$ denotes the value of the Ricci scalar at present.
Also, for a de Sitter attractor point where the Ricci scalar takes the value $R_1>0$, then
$f'(R)>0$ must be satisfied also for $R>R_1>0$. This is a way to avoid antigravity regimes.
  \item $f''(R)>0$ for $R>R_0$, in order for cosmological perturbation to be stable,
and for the consistency with local gravity tests during the matter-dominated epoch.
  \item $f(R)\rightarrow R-\Lambda$ for $R\rightarrow\infty$, in order the model is reduced to the $\mathrm{\Lambda}\mathrm{CDM}$ model at large curvatures and in order inflation is ensured.
  \item $m^2=\frac{1}{3}\left(\frac{f'(R)}{f''(R)}-R\right)>0$, to avoid having tachyonic
instability of the scalaron in the Einstein frame equivalent theory of the Jordan frame $f(R)$ theory.
  \item $f(0)=0$ for $R=0$, to ensure that a Minkowski spacetime solution exists.
\end{itemize}
Most of the models we study in this section satisfy the majority (or all) the aforementioned fitness tests and the parameters values we use are chosen in such a way that the above fitness requirements are fulfilled. In the following subsections we present the numerical study we performed for $f(R)$ models with or without collisional matter. In addition, all the results shall be directly compared to the $\Lambda\mathrm{CDM}$ model.

\subsection{Study of Some Realistic Exponential $f(R)$ models and Comparison to the $\Lambda\mathrm{CDM}$ Model}

In this subsection we examine various exponential models that satisfy all of the constraints that render an $f(R)$ theory viable. Particularly, we critically investigate what is the effect of collisional matter in exponential $f(R)$ theories. Exponential $f(R)$ models of gravity were studied for the first time in \cite{importantpapers11} by Cognola, Elizalde et al., but see also the work of Bamba et al. \cite{importantpapers12} and  that of Linder \cite{exponentialcosmo2} for interesting and thorough studies on exponential models in $f(R)$ theories. 
\begin{figure}[h]
\centering
\includegraphics[width=15pc]{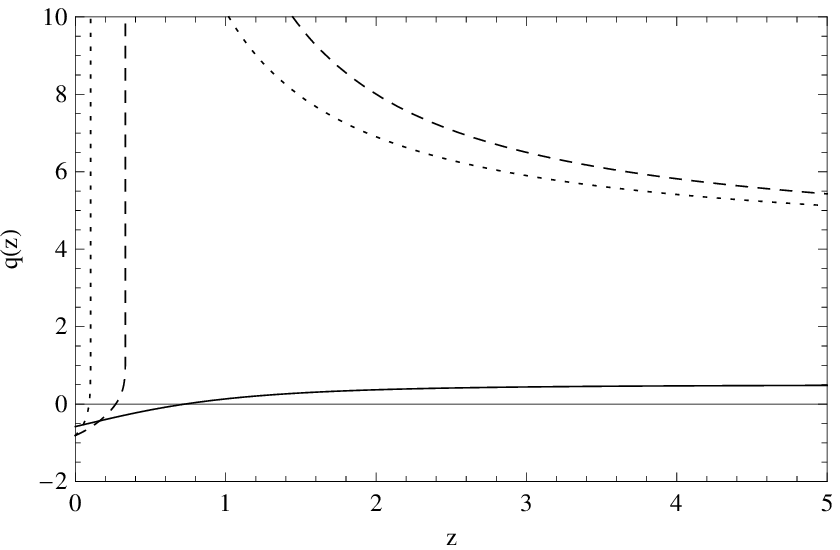}
\includegraphics[width=15pc]{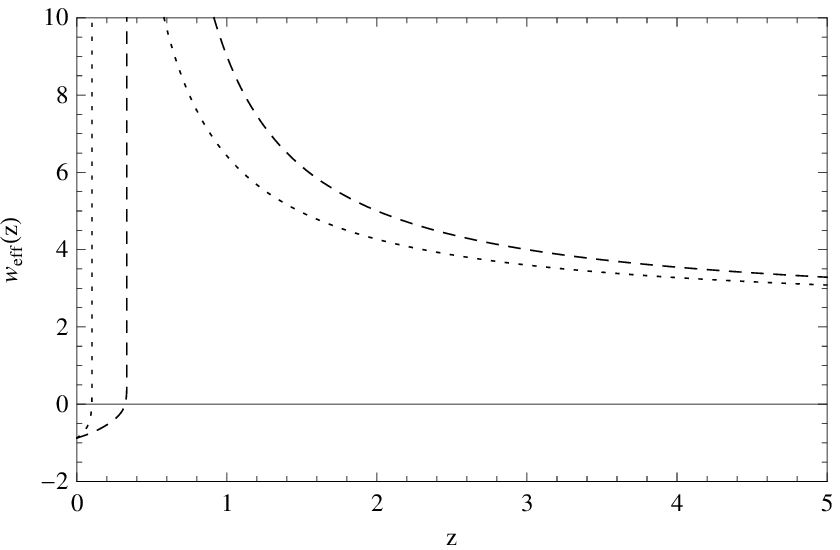}
\caption{Plots of $q(z)$ and $w_{eff}$
over z, for the model
$f(R)=R-bR_s(1-e^{-R/R_s})$. The dashed, dotted, solid lines refer to collision-less, collisional, and
$\Lambda\mathrm{CDM}$ models, respectively.}\label{fig:elizaldeplots1}
\end{figure}
We start our investigation with one of the most frequently used and much appealing exponential $f(R)$ model of reference \cite{importantpapers11}, since this model passes all viability test and more importantly does not lead to finite time singularities. The functional form of this model is of the following form:
\begin{equation}\label{eimod}
  f(R)=R-bR_s(1-e^{-R/R_s}),
\end{equation}
where the parameters in this example are chosen to be those appearing in reference \cite{importantpapers12}. We have performed the same analysis as before and in Fig. \ref{fig:elizaldeplots1} we have plotted the results. As in the previous cases, the left plot depicts the $z$-dependence of the deceleration parameter $q(z)$, while the right plot the $z$ dependence of the effective equation of state parameter $w_{eff}$ . As we can see from the left plot of Fig. \ref{fig:elizaldeplots1}, as in the previous case, the curves of the collisional matter $f(R)$ theory (dotted) and that of the pressure-less $f(R)$ theory (dashed) are quite close to each other, with the transition redshift corresponding to the collisional matter $f(R)$ theory case, being slightly smaller in comparison to the pressure-less matter $f(R)$ theory and $\Lambda\mathrm{CDM}$ ones. In addition, we observe the same behavior for the collisional matter $f(R)$ theory (hereafter C-$f(R)$) and the pressure-less matter $f(R)$ theory (hereafter P-$f(R)$), for both the $q(z)$ and $w_{eff}$. In reference to the $q(z)$, both $f(R)$ theories have the deceleration-acceleration transition, with the transition for the C-$f(R)$ theory occurring at smaller redshifts, in comparison to the P-$f(R)$ case. Moreover, both curves have a pick around $z=1$ which smooths out after $z\simeq 1.8$. Such a behavior indicates that for the $f(R)$ model (\ref{eimod}) the universe decelerated at high $z$, with deceleration that gradually increased, with the lowest deceleration occurring at $z\simeq 1$, and after that, gradually the deceleration started to decrease and at some point below $z=1$, the universe started to accelerate. By looking the $w_{eff}$ plot we can see that in both the C-$f(R)$ and P-$f(R)$ models, the effective equation of state parameter $w_{eff}$ does not cross the phantom divide $w=-1$ \cite{phantom}. Let us note here for clarity, that the value $w=-1$, as is known from the literature \cite{reviews1},  corresponds to a cosmological constant, and if $-1<w<0$, we have quintessence energy, while when $w<-1$ we have the phantom energy region. In addition, observations of supernovas, mainly from SNe Ia ones \cite{phantom}, seem to favor an equation of state for dark energy which has crossed the phantom divide $w=-1$ in the near past. This behavior however cannot be seen in Fig. \ref{fig:elizaldeplots1}. Another notable feature of both the $q(z)$ and $w_{eff}$ plots is that in both cases, the curves asymptotically approach each other, something that occurs for $z>4$ in both plots. As a final comment, regarding the model (\ref{eimod}), if we lower the value of $b$ and raise the value of $R_s$ then the transition redshift $z_t$ acquires a lower value and, at the same time, the plateau at which both curves tend to go asymptotically for large $z$ becomes higher, referring to the right plot of Fig. \ref{fig:elizaldeplots1}.
\begin{figure}[h]
\centering
\includegraphics[width=15pc]{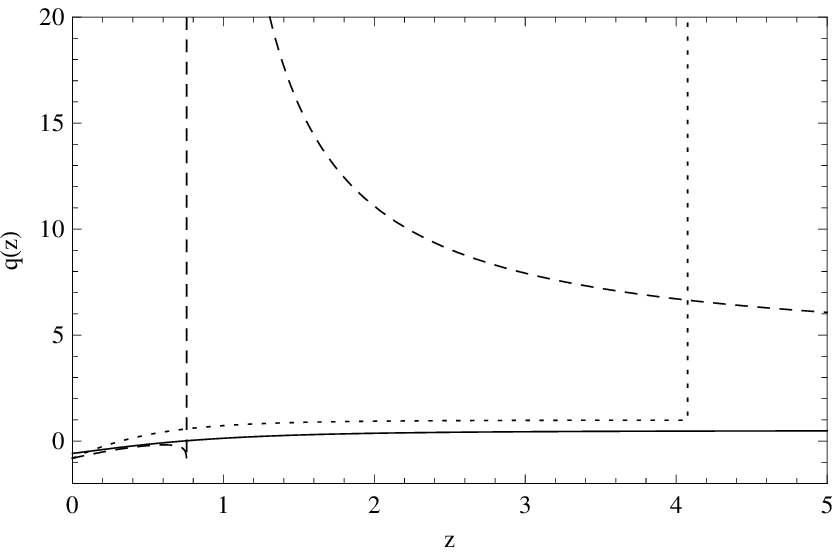}
\includegraphics[width=15pc]{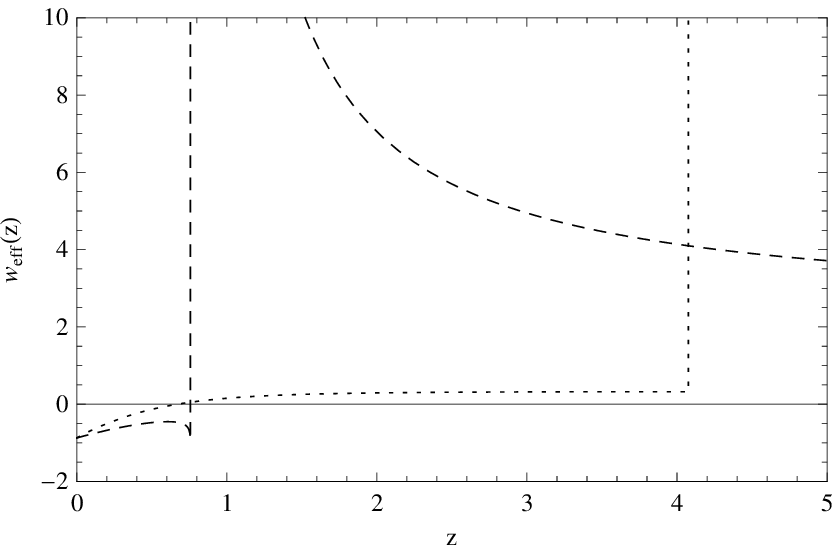}
\caption{Plots of $q(z)$ and $w_{eff}$ over z, for the model
$f(R)=R-\frac{C}{A+Be^{-R/D}}+\frac{C}{A+B}$. The dashed, dotted, solid lines refer to
collision-less, collisional, and $\Lambda\mathrm{CDM}$ models, respectively.}\label{fig:oikonomouplots1}
\end{figure}
The second exponential model we shall study is a viable exponential model studied in reference \cite{oikonomoupaper}. The functional form of the model is described by the following:
\begin{equation}\label{oikonomod}
  f(R)=R-\frac{C}{A+Be^{-R/D}}+\frac{C}{A+B},
\end{equation}
with $A$, $B$, $C$, and $D$ being constant parameters. For stability reasons these parameters have
to respect the following relations \cite{oikonomoupaper} :
\begin{equation}
  A>B \text{ and } D>C,
\end{equation}
For the numerical analysis we performed, we chose the parameters values to be similar to those of reference \cite{oikonomoupaper}. The results of the numerical analysis we performed can be found in Fig. \ref{fig:oikonomouplots1}. This case shows notable features which we now discuss in detail. Again in the plots, the dashed, dotted, solid lines refer to
P-$f(R)$, C-$f(R)$, and $\Lambda\mathrm{CDM}$ models, respectively. As we can see in the $q(z)$ plot, the C-$f(R)$ model curve is almost identical to the $\Lambda\mathrm{CDM}$ curve and this occurs until the redshift $z=4$, at which redshift, $q(z)$ for C-$f(R)$ shows a pick that smooths out after $z=6$ (not shown in the plot). In addition, the transition redshift for the C-$f(R)$ and for the $\Lambda\mathrm{CDM}$ models, have almost the same value. As for the deceleration-acceleration behavior, in both the C-$f(R)$ and P-$f(R)$ models, the universe decelerated at high $z$, with deceleration that gradually increased (for the P-$f(R)$ model around $z=2.5$ and for the C-$f(R)$ model around $z=6.5$), with the lowest deceleration at $z\simeq 5$ for the C-$f(R)$ and at $z\simeq z=1.5$ for the P-$f(R)$. After that, in both cases the universe's deceleration gradually  started to decrease and at the corresponding transition redshift, the universe started to accelerate. The same applies for the $w_{eff}$ plot, so we refrain from going into details. Let us note once more the resemblance of the $\Lambda\mathrm{CDM}$  and the C-$f(R)$ curves.

\subsection{Study of Power-law $f(R)$ Models with and without Collisional Matter and Comparison to $\Lambda\mathrm{CDM}$}

\begin{figure}[h]
\centering
\includegraphics[width=15pc]{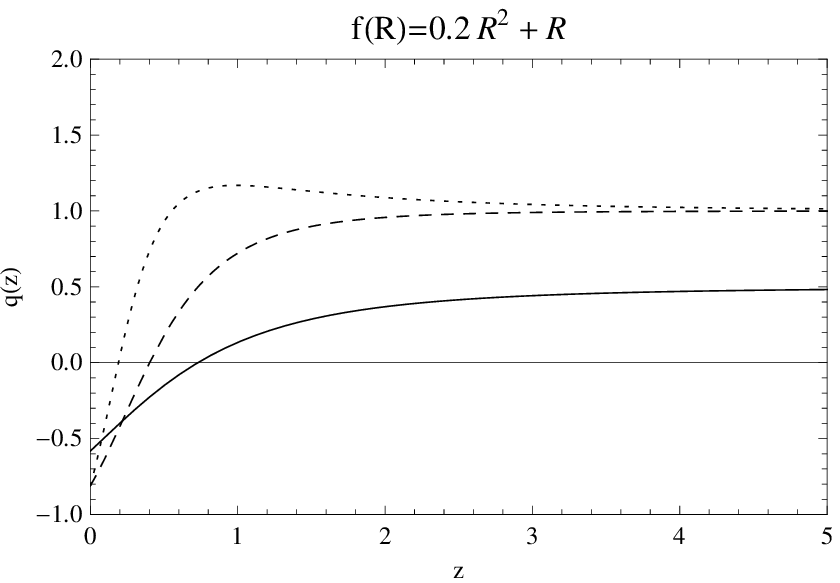}
\includegraphics[width=15pc]{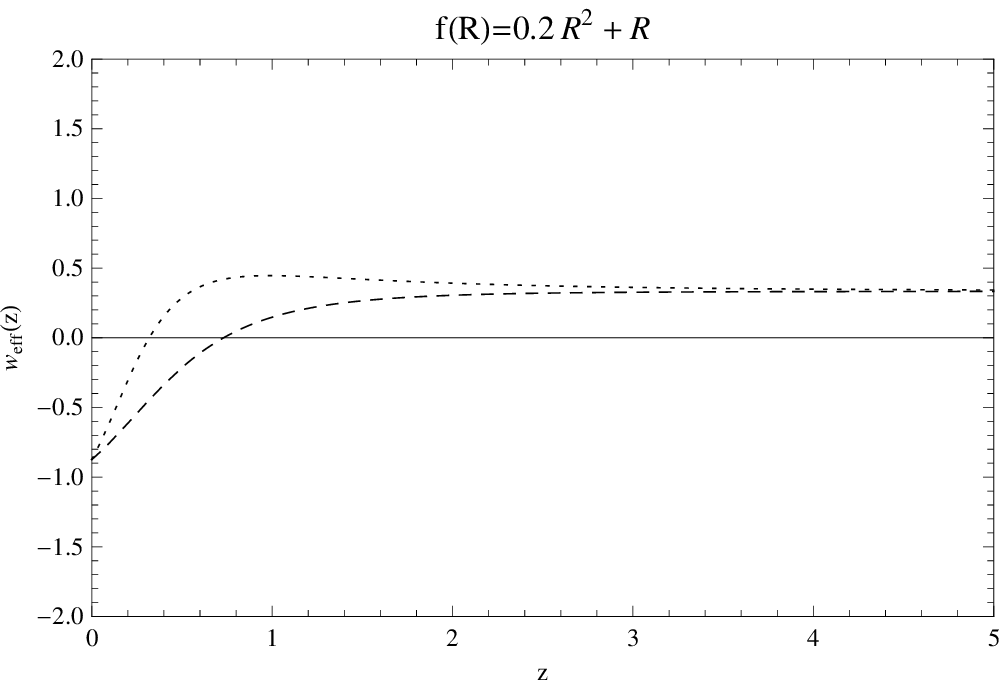}
\caption{Plots of $q(z)$ and $w_{eff}$ over z, for the model
$f(R)=R+aR^2$. The dashed, dotted,
solid lines refer to collision-less, collisional, and
$\Lambda\mathrm{CDM}$ models, respectively.}\label{fig:powerlaw1plots1}
\end{figure}

The last category of known viable $f(R)$ models we shall work on is the power-law models \cite{reviews1,reviews101,reviews2}. The first model we shall work with is quite well known for it's inflationary solutions \cite{reviews1,reviews101,reviews2} (actually the $f(R)\sim R^2$ model), with the general functional form being the following:
\begin{equation}\label{powerlawinfl}
  f(R)=R+aR^n
\end{equation}
We focus on the case with $n=2$, since this is the most well known form of this modified gravity model, with $a=0.02$. After performing the general numerical analysis in the same way as in the previously studied cases, we found similar results which we presented in Fig. \ref{fig:powerlaw1plots1}. The plots corresponding to this model are quite smooth and describe a mild late time evolution of the universe, with a smooth transition from deceleration to acceleration. The curve of the P-$f(R)$ model is closer to the $\Lambda\mathrm{CDM}$ model one, with the C-$f(R)$ curve giving a lower transition redshift in comparison to the other two models.
With regards to the $w_{eff}$ plot, both the P-$f(R)$ and the C-$f(R)$ model show similar behavior and both curves do not cross the phantom divide. 

\begin{figure}[h]
\centering
\includegraphics[width=15pc]{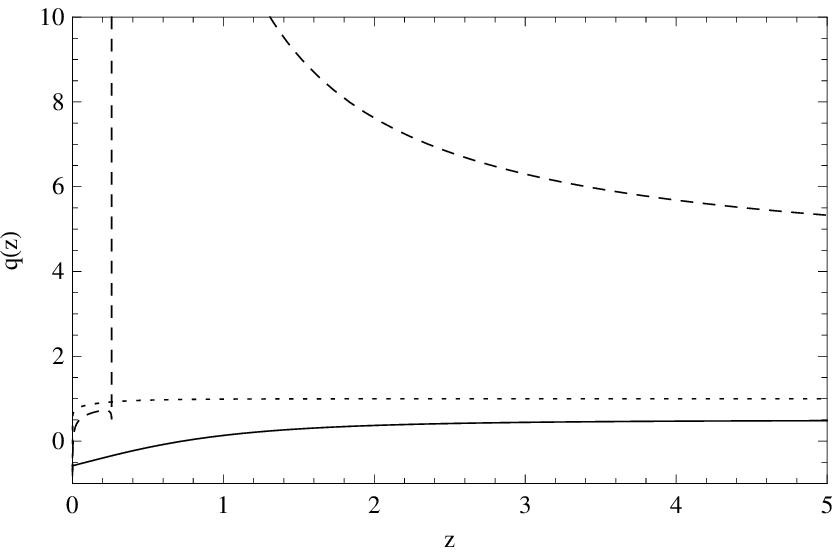}
\includegraphics[width=15pc]{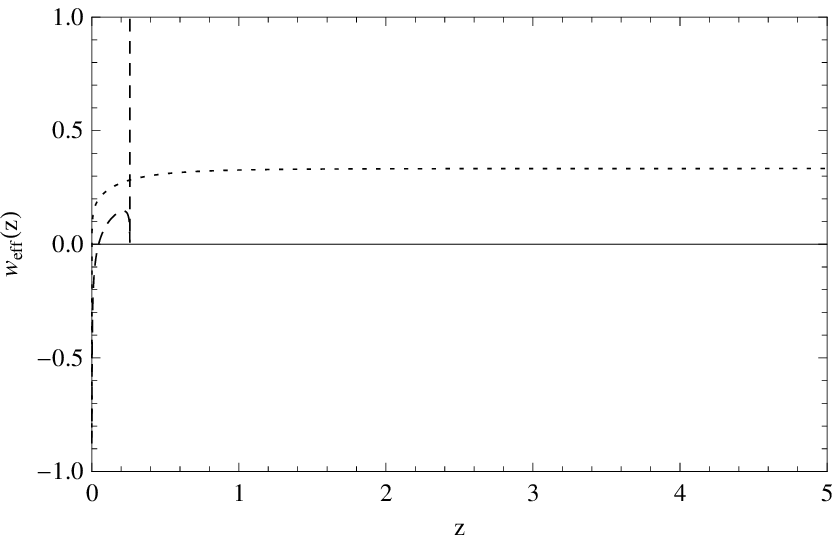}
\caption{Plots of $q(z)$ and $w_{eff}$ over z, for the model
$f(R)=R+aR^{-2}$. The dashed, dotted,
solid lines refer to collision-less, collisional, and
$\Lambda\mathrm{CDM}$ models, respectively.}\label{fig:powerlaw1plots2}
\end{figure}

\begin{figure}[h]
\centering
\includegraphics[width=15pc]{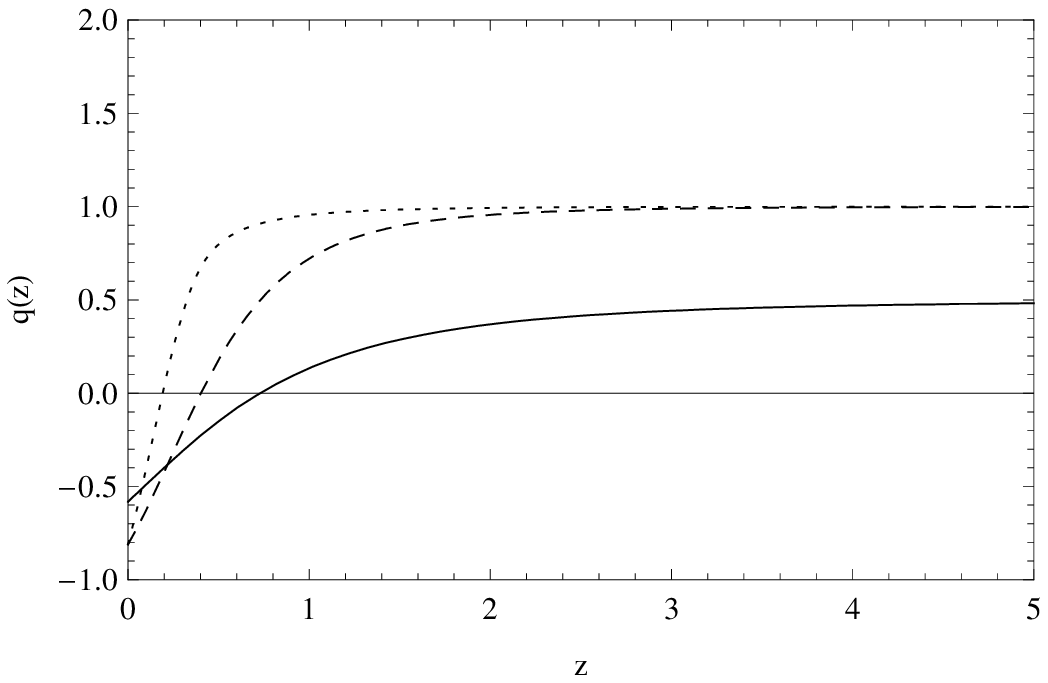}
\includegraphics[width=15pc]{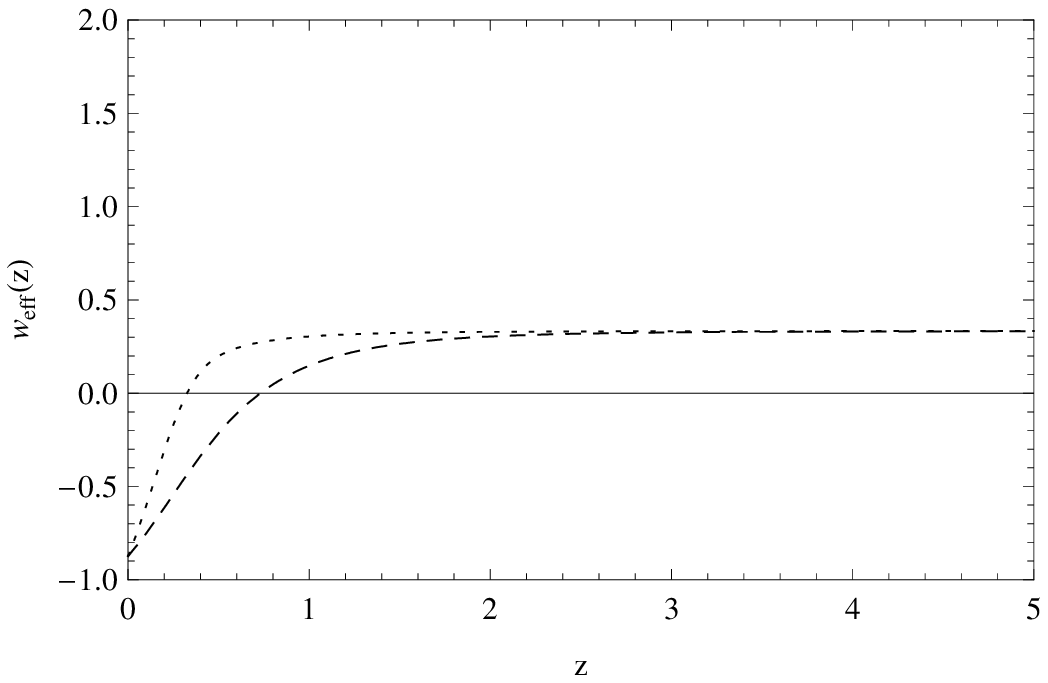}
\caption{Plots of $q(z)$ and $w_{eff}$ over z, for the model
$f(R)=R+\alpha R^{2}-\frac{\beta}{R^n}$. The dashed, dotted,
solid lines refer to collision-less, collisional, and
$\Lambda\mathrm{CDM}$ models, respectively.}\label{fig:powerlaw1plotssergei}
\end{figure}

Another very well known power-law model we shall present is described by the following functional form:
\begin{equation}\label{sasy}
  f(R)=R+aR^{-n},
\end{equation}
We performed the same numerical analysis as in all the previous cases and we presented the results in Fig. \ref{fig:powerlaw1plots2}, for $n=2$.  Without going into details about the results, we shall discuss only one intriguing feature of this model. Noticing the C-$f(R)$ curve in both $q(z)$ and $w_{eff}$, we can see that in both plots it is particularly smooth. Specifically, in the $q(z)$ plot, the C-$f(R)$ is very close to the $\Lambda\mathrm{CDM}$. This feature seems to be very model dependent and we came across with such a behavior in the study of the exponential model (\ref{oikonomod}), see Fig. \ref{fig:oikonomouplots1}. All the rest analysis for the numerical results of this model is similar to the previous models analysis, so we refrain from going into details.
The last power law model we shall present is described by the following $f(R)$ function,
\begin{equation}\label{fsergeinojiri}
f(R)=R+\beta R^2-\frac{\alpha}{R^n}
\end{equation}
and was introduced for the first time in \cite{sergeinojirimodel}, as a model that offers a consistent description of late time acceleration and inflation. Performing the same numerical analysis as in the previous cases, in Fig. (\ref{fig:powerlaw1plotssergei}) we have presented the late time behavior of the deceleration parameter (left) and of the effective equation of state parameter $w_{eff}$, as functions of the redshift $z$. The numerical values for the parameters are taken to be identical with the ones used in \cite{sergeinojirimodel} and also we used $n=3$.
As we can see, the late time behavior of the C-$f(R)$ theory differs from the P-$f(R)$ theory and this can be seen in both the $q(z)$ and $w_{eff}$ plots. Specifically, the P-$f(R)$ theory gives better fitting to the $\mathrm{\Lambda}\mathrm{CDM}$ model, with the C-$f(R)$ theory giving a transition redshift smaller in comparison to the ones given by the $\mathrm{\Lambda}\mathrm{CDM}$ and the P-$f(R)$ models. In the next section we shall see in a quantitative way why the collisional matter $f(R)$ theory differs from the pressure-less matter $f(R)$ model.

\section{Possibility of Connecting Matter Dominating Epoch with Dark Energy Universe}

In the previous sections we mainly focused on the late time cosmological evolution of the universe in $f(R)$ theories, with the additional contribution of collisional matter. However, it is necessary to study whether the effect of collisional matter actually has to offer something new to the important issue of connecting the matter dominating epoch with the late time acceleration epoch. A quite thorough and detailed study on this issue was performed in reference \cite{importantpapers3}. We shall adopt the notation and their approach since, as we shall demonstrate, the only viable possibility of connecting the matter dominating epoch with the present time acceleration is to use the concept of compensating dark energy, introduced for the first time in \cite{importantpapers3} by Nojiri and Odintsov. In order to have concordance to the notation of \cite{importantpapers3}, we alter the definition of the metric we used in the previous sections, namely equation (\ref{metricformfrwhjkh}) by reversing the signature, so that the new metric is:
\begin{equation}\label{metricformfrwhjkhgfgf}
\mathrm{d}s^2=\mathrm{d}t^2-a^2(t)\sum_i\mathrm{d}x_i^2
\end{equation}
and therefore the Ricci scalar in this background reads:
\begin{equation}\label{ricciscafgfgjfghl}
R=6(2H^2+\dot{H}),
\end{equation}
The novel approach of reference \cite{importantpapers3} is that the matter dominated and late time acceleration epoch was directly derived from the $f(R)$ theory. In our case however, collisional matter effects cannot be derived from the $f(R)$ theory directly so the effect of matter has to be introduced by hand. Hence, we have no other choice but introducing matter and collisional matter by hand from the beginning.

We shall investigate the $f(R)$ model appearing in equation (\ref{fsergeinojiri}) which was studied in \cite{importantpapers3,sergeinojirimodel}, and as was demonstrated, it is able to predict the existence of inflationary, matter dominating and late time acceleration eras in the universe (with $n$ taken to be equal to some positive integer). Recall that the late time behavior of this model was studied in the end of the previous section. The total mass-energy density $\rho_{tot}(t)$ and the total pressure $p_{tot}(t)$ of the incompressible fluids described by the $f(R)$ theory and also from the matter and collisional matter, are equal to:
\begin{align}\label{colenergymatt1}
& \rho_{tot}(t) =f(R)-6\Big{(}\dot{H}+H^2-H\frac{\mathrm{d}}{\mathrm{d}t}\Big{)}f'(R)+\varepsilon_m \\ \notag &
p_{tot}(t) =f(R)-6\Big{(}\dot{H}+H^2-H\frac{\mathrm{d}}{\mathrm{d}t}\Big{)}f'(R)+w\varepsilon_m
\end{align}
with $\varepsilon_m$ defined in (\ref{totenergydens}). We make the same assumption for the Hubble parameter $H(t)$ made in \cite{importantpapers3}, that is:
\begin{equation}\label{hubass1}
H(t)=\frac{h(t)}{t}
\end{equation}
with $h(t)$ a slowly varying function (which means that it's derivatives can be neglected). The latter attribute of the function $h(t)$ is of crucial importance for the rest of this section. In addition, we assume that the scale factor $a(t)$ varies as follows:
\begin{equation}\label{scalefucntvariation}
a(t)=a_0e^{g(t)}
\end{equation}
with $g(t)$,
\begin{equation}\label{gtsddff}
g(t)=h(t)\ln (t/t_i)
\end{equation}
and $t_i$ some initial time. Finally, we suppose that the matter-energy density $\varepsilon_m$ varies as a function of the scale factor as in relation (\ref{evlotuniomatt}). Therefore, by using relations (\ref{colenergymatt1}), (\ref{hubass1}), (\ref{scalefucntvariation}) and (\ref{gtsddff}), we may study the behavior of the total energy density $\rho_{tot}(t)$ and directly compare it to the various curvature terms appearing in (\ref{fsergeinojiri}). We start off by examining the inflationary era, which is described by the term $f(R)\sim \beta R^2$ and occurs say at $t_0$. By using (\ref{hubass1}), the total energy density $\rho_{tot}$ is (we only present the energy density for simplicity, the total pressure can also easily be found) \cite{importantpapers3}:
\begin{align}\label{inflerarho}
& \rho_{tot}(t)=\rho_c(t)+\rho_M(t)+\rho_{col}(t)\sim -\frac{36\beta \Big{(}-1+2h(t)\Big{)}h(t)^2}{t^{4n}}\\ \notag &
+\rho_{m0}e^{-3\int_{t_0}^t\frac{h(t)}{t}\mathrm{d}t}(\Pi_0+1)-3\rho_{m0}e^{-3\int_{t_0}^t\frac{h(t)}{t}\mathrm{d}t}\Big{(}-3\int_{t_0}^t\frac{h(t)}{t}\mathrm{d}t\Big{)}
\end{align}
where with $\rho_c$ we denote the curvatures energy density contribution (which is of geometric origin) and with $\rho_M$ and $\rho_{col}$ the ordinary matter and purely collisional matter contributions, which are:
\begin{align}\label{inflerarhostartengines}
& \rho_M(t)= \rho_{m0}e^{-3\int_{t_0}^t\frac{h(t)}{t}\mathrm{d}t}\\ \notag &
\rho_{col}\Pi_0\rho_{m0}e^{-3\int_{t_0}^t\frac{h(t)}{t}\mathrm{d}t}-3\rho_{m0}e^{-3\int_{t_0}^t\frac{h(t)}{t}\mathrm{d}t}\Big{(}-3\int_{t_0}^t\frac{h(t)}{t}\mathrm{d}t\Big{)}
\end{align}
The last two terms of the above equation (\ref{inflerarho}) contain the contribution of the collisional matter and particularly the following two:
\begin{align}\label{twotermscollisionalmatter}
\Pi_0\rho_{m0}e^{-3\int_{t_0}^t\frac{h(t)}{t}\mathrm{d}t},{\,}{\,}{\,}-3\rho_{m0}e^{-3\int_{t_0}^t\frac{h(t)}{t}\mathrm{d}t}\Big{(}-3\int_{t_0}^t\frac{h(t)}{t}\mathrm{d}t\Big{)}
\end{align}
Notice the first term proportional to $\Pi_0$ behaves exactly like the ordinary matter does, so it is senseless to study the behavior of the total energy density without ordinary matter, there is practically no difference. Hence, the problem at hand is reduced to just comparing the total energy density (\ref{inflerarho}) with the inflation generating term $f(R)\sim R^2$, which in terms of the slowly varying function $h(t)$ behaves as \cite{importantpapers3}:
\begin{equation}\label{uinflfyutre}
f(R)\sim \beta R^2\sim h^8 
\end{equation}
Note that in reference \cite{importantpapers3}, the approximation (\ref{uinflfyutre}) holds true only when $h\rightarrow \infty$, which corresponds to a de Sitter universe. This is quite logical, since the $t\ll 1$, when $t\sim t_0$. As was explained in \cite{importantpapers3}, as $h(t)$ tends to infinity, which is actually the de Sitter universe, the first term of relation (\ref{inflerarho}) behaves as $\sim h^3$. Now what remains is to see how the rest terms of (\ref{inflerarho}) behave. In order to do so, we have to use some analysis theorems and define in a formal way the general features of the function $h(t)$. Since the universe is evolving in an expanding way, the function $H(t)$ is a monotonically decreasing function of time which. It is easy to see this, since $H(t)=h(t)/t$, and, as already stated, $h(t)$ varies slowly with time. Actually in reference \cite{importantpapers3} one choice for $h(t)$ was the following (see equation (44) of reference \cite{importantpapers3}):
\begin{equation}\label{sergeischoci}
h(t)=\frac{h_i+h_fqt^2}{1+qt^2}
\end{equation}
which has a positive first derivative owing to the fact that $h_f>h_i$ (that is $h(t)$ is monotonically increasing as a function of $t$). In addition, with the choice (\ref{sergeischoci}), the function $H(t)$ is made a monotonically decreasing function of time, a fact that proves to be crucial to our analysis. Assuming such monotonicity properties for $H(t)$ and $h(t)$, we proceed to some definitions. Slowly varying functions have been thoroughly studied in the literature \cite{slowlyvar} and we now briefly present the issues we shall need for our study. It is known that \cite{slowlyvar} if $h(t)$ is a slowly varying function, it can actually take the following form:
\begin{equation}\label{slowlvar1}
h(t)=c(t)e^{\int_{t_0}^t\frac{y(x)}{x}}\mathrm{d}x
\end{equation}
with $c(t)$ a measurable non-negative function of $t$, and $c(t)$ and $y(x)$ are defined in such a way, so that the following requirements are met:
\begin{equation}\label{requirem1}
\lim_{t\rightarrow \infty}c(t)=c_0,{\,}{\,}{\,}\lim_{x\rightarrow \infty }y(x)\rightarrow 0
\end{equation}
with $c_0$ a finite number. In addition, owing to the monotonicity of $h(t)$, the functions $c(t)$ and $y(x)$ are defined accordingly. Now we make use of the fact that $h(t)$ is monotonically increasing as a function of time and also that $h(t)/t$ is monotonically decreasing and define the quantity:
\begin{equation}\label{newquant}
B(t)=\rho_{m0}e^{-3\int_{t_i}^t\frac{h(t)}{t}\mathrm{d}t}(\Pi_0+1)-3\rho_{m0}e^{-3\int_{t_i}^t\frac{h(t)}{t}\mathrm{d}t}\Big{(}-3\int_{t_i}^t\frac{h(t)}{t}\mathrm{d}t\Big{)}
\end{equation}
Owing to the exponential time dependence of the above expression, the maximum value $B_{t}$, which we denote $B_{max}$ can be achieved in two cases, either when $t\rightarrow \infty$, or when $t\simeq t_{min}$ with $t_i\leq t_{min} \leq t_0$, depending on the choice of the function $h(t)$. The extreme case $t_{min}\rightarrow \infty $ is rather a difficult case that involves a time which is beyond the time interval of our interest. But in this case too, we can get some results for the behavior of $B(t)$ as we will see later on where we explain the conditions under which this may occur. Lets focus on the choice $t_{min}\neq \infty$ and $t_i\leq t_{min} \leq t_0$. In this case, the exponential receives the maximum value when the exponent is the smallest, which is achieved at a time $t_i\leq t_{min} \leq t_0$, at which time the function $\int_{t_i}^{t}\frac{h(t)}{t}\mathrm{d}t$ has a global minimum. Notice that for the choice (\ref{sergeischoci}), the integral reads:
\begin{equation}\label{intreads}
\int_{t_i}^t\frac{h(t)}{t}\mathrm{d}t=h_i \ln t\Big{|}_{t_i}^{t}+\frac{1}{2}(h_f-h_i)\ln \Big{(}1+q t^2\Big{ )}\Big{|}_{t_i}^{t}
\end{equation}
so $B_{max}$ is achieved when $t \simeq t_0$. At this point it is crucial to make the assumption that the function $\int_{t_i}^{t}\frac{h(t)}{t}\mathrm{d}t$ takes positive values in the interval $[t_i,t_0]$. The latter constraint is consistent with the assumption made in reference \cite{importantpapers3}, that is, $h(t)\rightarrow \infty$ and also that the choice (\ref{sergeischoci}) made in reference \cite{importantpapers3}, along with all other choices made in that work, actually respect the condition $\int_{t_i}^{t}\frac{h(t)}{t}\mathrm{d}t>0$, for $t$ $\in$ $[t_i,t_0]$. Note that if $\int_{t_i}^{t}\frac{h(t)}{t}\mathrm{d}t<0$, the addition of collisional matter completely dominates the inflationary era and therefore the $f(R)\sim R^2$ term does not dominate anymore. Therefore, the inflationary picture can be fatally harmed for the choice (\ref{sergeischoci}) and for the model (\ref{fsergeinojiri}). This however never happens for all the reasonable choices of $h(t)$ made in \cite{importantpapers3}, since $t_0,t_i\ll 1$ during the inflationary era. Nevertheless, this result is strongly model dependent and also depends on the choice of the slowly varying function $h(t)$. In addition, since the additional ordinary matter contribution 
\begin{equation}\label{tispoutanas}
\rho_M(t)=\rho_{m0}e^{-3\int_{t_0}^t\frac{h(t)}{t}\mathrm{d}t}
\end{equation}
harms inflation in the same way, if $\int_{t_i}^{t}\frac{h(t)}{t}\mathrm{d}t<0$, for $t$ $\in $ $[t_i,t_0]$. Therefore, the addition of collisional matter at the inflation stage is as harmful as ordinary matter is. But the constraint $\int_{t_i}^{t}\frac{h(t)}{t}\mathrm{d}t<0$ is respected for all physical choices of the function $h(t)$ and hence the fatal condition $\int_{t_i}^{t}\frac{h(t)}{t}\mathrm{d}t<0$ never occurs actually.

Assuming that $h(t)$ is chosen in a such a way so that $\int_{t_i}^{t}\frac{h(t)}{t}\mathrm{d}t>0$ for any $t$ $\in$ $[t_i,t_0]$, for a generic $t_{min}$ value in the interval $[t_i,t_0]$, the maximum value of $B(t)$, that is $B_{max}$, is given by:
\begin{align}\label{newquantmax1}
& B_{max}=\lim_{t\rightarrow t_{min}}B(t)=\mathrm{sup}\Big{(}\rho_{m0}e^{-3\int_{t_0}^t\frac{h(t)}{t}\mathrm{d}t}(\Pi_0+1)-3\rho_{m0}e^{-3\int_{t_0}^t\frac{h(t)}{t}\mathrm{d}t}\Big{(}-3\int_{t_0}^{t}\frac{h(t)}{t}\mathrm{d}t\Big{)}\Big{)} \\ \notag &
=\Big{(}\rho_{m0}e^{-3\int_{t_0}^{t_{min}}\frac{h(t)}{t}\mathrm{d}t}(\Pi_0+1)-3\rho_{m0}e^{-3\int_{t_0}^{t_{min}}\frac{h(t)}{t}\mathrm{d}t}\Big{(}-3\int_{t_0}^{t_{min}}\frac{h(t)}{t}\mathrm{d}t\Big{)}\Big{)}
\end{align}
and since $B(t)\leq B_{max}$ for any $t_i\leq t_{min} \leq t_0$, then we can compare $B_{max}$ with $\rho_c(t_0)$ and find how the $B(t)$ behaves in comparison to $\rho_c(t_0)$. Indeed, if $B(t)\leq B_{max}$ for any $t$ with $t_i\leq t\leq t_0$, we have:
\begin{equation}\label{trela}
\rho_c(t_0)+B(t_0)\leq \rho_c(t_0)+B(t_{min})
\end{equation} 
with $B(t_{min})=B_{max}$. Any result holding true for $t=t_{min}$, will hold true for any time in the interval $[t_i,t_0]/\{t_{min}\}$, and hence for $t=t_0$ too. Now $B(t_{min})$ contains exponentials which go to zero sufficiently faster than the powers of $h(t)$, and therefore we get that the total energy density (\ref{inflerarho}) is given by:
\begin{align}\label{inflerarhoqw2q}
\rho_{tot}(t)\sim -\frac{36\beta \Big{(}-1+2h(t)\Big{)}h(t)^2}{t^{4n}}
\end{align}
which as was explained below equation (\ref{uinflfyutre}), behaves as $\rho_{tot}\sim h^3$ and clearly the curvature term given in equation (\ref{uinflfyutre}) dominates. Now let us discuss the case in which $t_{min}\sim \infty$. This can be true when the function $\int_{t_i}^{t}\frac{h(t)}{t}\mathrm{d}t$ is a monotonically increasing function of $t$. In this case, $B_{max}$ reads,
\begin{align}\label{newquant}
& B_{max}=\lim_{t\rightarrow \infty}B(t) \\ \notag &
=\Big{(}\rho_{m0}e^{-3\int_{t_0}^{\infty}\frac{h(t)}{t}\mathrm{d}t}(\Pi_0+1)-3\rho_{m0}e^{-3\int_{t_0}^{\infty}\frac{h(t)}{t}\mathrm{d}t}\Big{(}-3\int_{t_0}^{\infty}\frac{h(t)}{t}\mathrm{d}t\Big{)}\Big{)}
\end{align}
In order to proceed we shall make use of Karamata's theorem (\cite{slowlyvar}), which says that given a regularly varying function $f(x)$, we have:
\begin{align}\label{karamatatheorem}
&\int_{t_i}^{t}f(x)\mathrm{d}x \sim (a+1)^{-1}tf(t),{\,}{\,}{\,}\mathrm{for}{\,}{\,}{\,}a>-1,{\,}{\,}{\,}\mathrm{and}{\,}{\,}{\,}t\rightarrow \infty \\ \notag &
\int_{t_i}^{t}f(x)\mathrm{d}x \sim -(a+1)^{-1}tf(t),{\,}{\,}{\,}\mathrm{for}{\,}{\,}{\,}a<-1,{\,}{\,}{\,}\mathrm{and}{\,}{\,}{\,}t\rightarrow \infty
\end{align}
The number $a$ is called the index of the regularly varying function $f(x)$, which is defined as:
\begin{equation}\label{introvas}
\lim_{x\rightarrow \infty}\frac{f(\lambda x)}{f(x)}=\lambda^a
\end{equation}
It is known that every regularly varying function can be written in the following way:
\begin{equation}\label{fugoirecell}
f(x)=x^aL(x)
\end{equation}
with $a$ the index of the function $f(x)$, and $L(x)$ a slowly varying function, that is, a function satisfying the constraints (\ref{requirem1}) and in addition the following two:
\begin{equation}\label{djfvasvas}
\lim_{t\rightarrow \infty}t^{-\gamma}h(t)\rightarrow 0,{\,}{\,}{\,}\lim_{t\rightarrow \infty}t^{\gamma}h(t)\rightarrow \infty
\end{equation}
which holds true for every $\gamma >0$. Therefore we can state that the slowly varying function $h(t)$ is equal to $f(t)=h(t)t$, with $f(t)$ regularly varying function of $t$, with index $a=1$. Then the integral $\int_{t_i}^{\infty}h(x)/x\mathrm{d}x$ can be written in terms of the function $f(t)$,
\begin{equation}\label{intfewritt}
\int_{t_i}^{\infty}h(x)/x\mathrm{d}x=\int_{t_i}^{\infty}f(x)\mathrm{d}x
\end{equation}
Making use of Karamata's theorem (\ref{karamatatheorem}), we obtain:
\begin{equation}\label{intfewritt1}
\int_{t_i}^{t}f(x)\mathrm{d}x\sim 2f(t)t=h(t), {\,}{\,}{\,}\mathrm{as}{\,}{\,}{\,}t\rightarrow \infty
\end{equation}
in which we made use of the fact that the function $f(x)$ has index $1$. Thereby, the function $B_{max}$ can be approximated by the expression 
\begin{align}\label{newquant}
& B_{max}=\lim_{t \rightarrow \infty }B(t) \\ \notag &
=\Big{(}\rho_{m0}e^{-3\frac{h(t)}{t}}(\Pi_0+1)-3\rho_{m0}e^{-3\frac{h(t)}{t}}\Big{(}-3\frac{h(t)}{t}\Big{)}\Big{)}
\end{align}
in which case the exponentials are subdominant, in comparison to powers of $h(t)$ (assuming of course that $h(t)>0$, which is a physically acceptable requirement). Then, we may conclude that:
\begin{equation}\label{trelasasythanioseitypseis}
\rho_c(t_0)+B(t_0)\leq \rho_c(t_0)+B(\infty)\simeq \rho_c(t_0)
\end{equation}
and therefore the sum $\rho_c(t_0)+B(t_0)\sim \rho_c(t_0)$, in which case the total energy density may be approximated by $\rho_{tot}\sim \rho_c(t_0)\sim h^3$, which is dominated by the term $f(R) \sim R^2$. In conclusion, the inflationary era can be consistently described even if we take into account the contributions coming from collisional matter, under the general assumption that $\int_{t_i}^{t}\frac{h(t)}{t}\mathrm{d}t>0$, which is respected for any physically acceptable choice of the slowly varying function of $h(t)$. If this is not the case, then the addition of collisional matter can be as harmful as the addition of ordinary matter during the inflationary case, which is however highly unlikely to occur, since an unphysical choice of $h(t)$ would be required.

Next we proceed to the matter dominated epoch, in which case the term $f(R)\sim R$ dominates. Since during the matter epoch we expect that $h(t)\simeq 2/3$, the total matter-energy density becomes:
\begin{align}\label{inflerarhocarriematter}
& \rho_{tot}=\rho_c(t)+\rho_M(t)+\rho_{col}(t)\sim -\frac{32}{3t^2}+\rho_{m0}(1+\Pi_0)t^{-2}+\rho_{m0}t^{-4}
\end{align}
where as before with $\rho_c$ we denote the curvatures energy density contribution and with $\rho_M$ and $\rho_{col}$ the ordinary matter and purely collisional matter contributions, which for this case are equal to:
\begin{align}\label{inflerarhostartenginescarriematter}
& \rho_M(t)\simeq \rho_{m0}t^{-2}\\ \notag &
\rho_{col}\simeq  \rho_{m0}\Pi_0t^{-2}+\rho_{m0}t^{-4} 
\end{align}
Clearly, during the matter domination era, the geometric contribution of the energy-density, receives extra contributions form collisional and ordinary matter, with the leading terms in relation (\ref{inflerarhocarriematter}) the ones of order $\sim t^{-2}$. Therefore, the matter domination era is further enhanced with the addition of collisional matter.

Finally, at the late time acceleration era, which it is assumed that occurs after a time $t\geq t_1$, we expect that $h(t)=h_f$ and therefore (see \cite{importantpapers3} for details):
\begin{align}\label{inflerarhocarriematterlate}
& \rho_{tot}\sim t^{3(1+w_f)h_f}+\rho_{m0}(1+\Pi_0)t^{-3h_f}+\rho_{m0}t^{-6h_f}
\end{align}
with $2n=-(1+w_f)h_f$ and $w_f$ the equation of state parameter of geometric dark energy at late times \cite{importantpapers3}. Since the first term of (\ref{inflerarhocarriematterlate}) is dominating over the other two, we conclude that the addition of ordinary and collisional matter does not drastically modify the late time behavior of the model (\ref{fsergeinojiri}), at least in the context of the approximations we assumed to hold true. 
Actually this can be observed in Fig. (\ref{fig:powerlaw1plotssergei}), where the plots for the $f(R)$ theory containing ordinary pressure-less matter and the one containing collisional matter are presented. Note that the corresponding total energy densities for the collisional and collision-less $f(R)$ theories cases are actually equal to $\rho_c(t)+\rho_M(t)+\rho_{col}(t)$ and $\rho_c(t)+\rho_M(t)$. Therefore, the $\rho_{col}(t)$ term spoils to some extend the late time acceleration picture of the model under study. 

Although the extra contributions from ordinary and collisional matter are not dominant, their presence requires to introduce an extra component in the total energy density which will cancel their contribution and will provide us with the fully correct evolution of the universe, both at the matter and late time acceleration eras.  This is the subject of the following subsection.

\subsection{Dark Energy Compensate for Collisional Matter}

One the greatest challenges of modified gravity theories is to consistently address the problem of coincidence and also describing the smooth transition from a matter domination era to a late time acceleration. With regards to the latter, it is required that during the matter domination era there is enough matter-energy density so that galaxies and other stellar formation can be created. In addition, during the late time acceleration the matter energy density of ordinary matter must be enough reduced in order the late time acceleration can be described consistently. In the same vain the transition has to be described in a correct and smooth way. Although the model we described in the previous subsection can describe observations to some extend, and regardless the fact that the effects of collisional matter are non-leading during the inflation and late time acceleration era, they are present and in some way must be eliminated. Notice however that during the matter domination era the collisional matter and ordinary matter energy density is added to the curvature contribution, which is a plausible feature of the theory. Nevertheless, the dark energy contribution during the matter era has to be eliminated. In order these problems are solved in a formal way, Nojiri and Odintsov in \cite{importantpapers3} introduced the concept of compensating dark energy. The purpose of this section is to investigate whether the collisional matter contribution with energy density $\rho_{col}$ can act as compensating dark energy or if not, how can we generalize the dark energy compensate construction of \cite{importantpapers3}. As we will demonstrate, the collisional matter cannot act as dark energy compensate so that a generalization of the dark energy compensate is required, for a correct description of the universe's evolution.

Before getting started it worths describing in brief the dark energy compensate of \cite{importantpapers3}. Let $\rho_{d}(t)$ describe the universe's energy density at the matter era, which occurs at a time $t\geq t_{m}$ and also $\rho_{l}(t)$ the energy density of the universe at late times $t\geq t_1$. These are defined as follows \cite{importantpapers3}:
\begin{align}\label{rdcont}
& \rho_d(t)=\frac{32}{3t_m^2}e^{-3\int_{t_m}^{t}h(t)/t\mathrm{d}t} \\ \notag &
\rho_l(t)=\alpha\Big{(}6(n+1)(2n+1)h_f+6(n-2)h_f^2\Big{)}(-6h_f+12h_f^2)^{-n-1}t_1^{2n}e^{3(1+w_f)\int_{t_1}^th(t)/t\mathrm{d}t}
\end{align}
The dark energy compensate is defined to be:
\begin{equation}\label{determinat}
\rho_{R}(t)=\rho_c(t)-\rho_d(t)-\rho_l(t)
\end{equation}
with $\rho_c(t)$ acting as the total energy density of the model (\ref{fsergeinojiri}). We denote $\rho_c^{cor}(t)$, the correct energy density of the universe and with correct, we mean that it is the energy density with all the wanted features for a late time acceleration. Since at $t\sim t_m$, $\rho_c^{cor}(t_m)\sim \rho_d(t_m)$, the dark energy compensate near $t\sim t_m$ is approximately $\rho_{R}(t)\sim -p_{l}(t_m)$ \cite{importantpapers3}. Moreover, at $t\sim t_1$, the energy density is $\rho_c(t_1)\sim p_l$, so that the dark energy compensate takes the value $\rho_{R}(t_1)\sim -p_{d}(t_1)$. Consequently the dark energy condensate dominates at $t\simeq t_1$ and becomes smaller at late times \cite{importantpapers3}. Thus the dark energy compensate during the matter domination era, actually subtracts from the energy $\rho_c(t)$ the term $p_l(t)$ and during the late time acceleration era, subtracts the term $\rho_d(t)$. So finally we obtain a correct picture for the universe's evolution for the model (\ref{fsergeinojiri}). Note that with this construction, the matter domination late time acceleration eras are connected in an explicit way \cite{importantpapers3}. 

Let us investigate whether the collisional matter can act as a dark energy compensate. Recall that what is required is actually to have more mass during the matter era ($t\geq t_m$) and less mass during the late times ($t\geq t_1$). 
\begin{equation}\label{djvy}
\rho_{col}(t)\Pi_0\rho_{m0}e^{-3\int_{t_i}^t\frac{h(t)}{t}\mathrm{d}t}-3\rho_{m0}e^{-3\int_{t_m}^t\frac{h(t)}{t}\mathrm{d}t}\Big{(}-3\int_{t_0}^t\frac{h(t)}{t}\mathrm{d}t\Big{)}
\end{equation}
with $t_i$ appropriately chosen. During the matter domination era $t\geq t_m$, the collisional matter contribution is:
\begin{equation}\label{tcolmatt}
\rho_{col}(t_m)\simeq \Pi_0\rho_{m0}t_m^{-2}+\rho_{m0}t_m^{-4}
\end{equation}
So the contribution of the collisional matter is actually very welcome during the matter domination era, since it increases the matter energy density. However the problem arises at late times, since at $t\sim t_1$ and at infinite time, the collisional energy density contribution is:
\begin{equation}\label{tcolmatt}
\rho_{col}(t_1)\simeq \Pi_0\rho_{m0}e^{-h_f}+\rho_{m0}3\rho_{m0}e^{-h_f}(3h_f)
\end{equation}
which spoils the late time acceleration picture. So the only way we can have a correct description of the universe's evolution is to generalize the dark energy compensate concept of reference \cite{importantpapers3}. We shall call it collisional dark energy compensate (CDEC hereafter) and we denote it $\rho_{RC}(t)$. In our case the CDEC, is defined as follows:
\begin{equation}\label{cdec}
\rho_{RC}(t)=\rho_{tot}^{cor}(t)-\rho_d(t)-\rho_l(t)-\rho_M(t)-\rho_{col}(t)
\end{equation}
with $\rho_{tot}^{cor}(t)$ the correct energy density of the universe (that is, the one with all the wanted features for a correct evolution). In addition, $\rho_d(t)$ and $\rho_l(t)$ are defined in relation (\ref{rdcont}) and also $\rho_M(t)$ and $\rho_{col}(t)$ are defined in relation (\ref{inflerarhostartengines}), where $t_0$ is to be replaced with an appropriate initial time. In addition, the following relation holds true:
\begin{equation}\label{rflasetotal}
\rho_{tot}(t)=\rho_{c}(t)+\rho_M(t)+\rho_{col}(t)
\end{equation}
with $\rho_{tot}(t)$, the non-corrected total energy density predicted for the model (\ref{fsergeinojiri}). The two total energy densities are related as follows:
\begin{equation}\label{fhhfdhepitelous}
\rho_{tot}^{cor}(t)=\rho_{tot}(t)+p_{CR}(t)
\end{equation}
Hence the CDEC actually corrects the flaws of the model, as these are seen in the behavior of the uncorrected total energy density $\rho_{tot}(t)$. During the matter domination era, at $t\geq t_m$, $\rho_{tot}^{cor}(t_m)\simeq \rho_d(t_m)$ and therefore the CDEC takes the value:
\begin{equation}\label{valueofcdec}
\rho_{RC}(t_m)\simeq -\rho_l(t_m)-\rho_M(t_m)-\rho_{col}(t_m)
\end{equation}
and at $t\geq t_1$, that is during the late time acceleration era, the correct total energy density is $\rho_{tot}^{cor}(t_1)\simeq \rho_l(t_1)$ and the CDEC value is equal to:
\begin{equation}\label{valueofcdec}
\rho_{RC}(t_1)\simeq -\rho_d(t_1)-\rho_M(t_1)-\rho_{col}(t_1)
\end{equation}
Thereby, we obtain the correct behavior for the total energy density of the universe for all times, since the flaws of the model, as these as seen at the function $\rho_{tot}(t)$, are corrected by the addition of the collisional dark energy compensate $\rho_{RC}(t)$. Notice that the CDEC dominates after $t\geq t_m$ and decreases at late times. 

Before closing this section, we shall give two additional definitions of the collisional dark energy compensate, that actually take into account the fact that the addition of collisional matter enhances the matter era, increasing the energy density of matter during that period of time. So the first variant definition of the one appearing in (\ref{cdec}) is the following:
\begin{equation}\label{cdecvar1}
\rho_{RC}(t)=\rho_{tot}^{cor}(t)-\rho_d(t)-\rho_l(t)-\Big{(}\rho_M(t)+\rho_{col}(t)\Big{)}\Theta (t-t_1)
\end{equation}
with $\Theta (t-t_1)$ the theta step function. In this way, the collisional matter contribution is added to the correct total energy density contribution so that at $t\simeq t_m$, we have $\rho_{tot}^{cor}(t_m)=\rho_d(t_m)+\rho_M(t_m)+\rho_{col}(t_m)$. The collisional dark energy at $t\geq t_1$ behaves as $\rho_{RC}(t_1)\simeq -\rho_d(t_1)-\rho_l(t_1)-\rho_M(t_1)-\rho_{col}(t_1)$, so that $\rho_{tot}^{cor}(t_1)\sim \rho_l(t_1)$. In this way, the matter domination epoch is enhanced by more matter energy density and the late time epoch remains unaffected. In addition, the transition between the two eras is ensured.

Another less appealing definition of the compensate dark is the following:
\begin{equation}\label{lessappealingcdedc}
\rho_{RC}(t)=\rho_{tot}^{cor}(t)-\rho_d(t)-\rho_l(t)-\int_{t_m}^t\Big{(}\rho_M(t)+\rho_{col}(t)\Big{)}\delta (t-t_1)\mathrm{d}t
\end{equation}
where we have used the following definition of the Dirac delta function:
\begin{equation}\label{dfjvhffhbyuf}
\int_c^df(t)\delta (t-a)\mathrm{d}t=f(a),{\,}{\,}{\,}\mathrm{if}{\,}{\,}{\,}c<a<d
\end{equation}
and the integral is zero unless $c<a<d$. Since the analysis of this form of the collisional dark energy compensate is similar to the one appearing in relation (\ref{cdecvar1}), we shall not go into details. Let us comment however that in the case described in relation (\ref{lessappealingcdedc}), a more smooth transition is achieved between the matter domination and late time acceleration eras.

\section*{Conclusions} 

The purpose of this article was to investigate the effect of collisional matter on the late time cosmological evolution of $f(R)$ theories of modified gravity. With the term collisional matter, it was meant that matter has some sort of self-interactions which we took into account. Collisional matter was considered to be a fluid with the $w$ parameter taking values $0<w<1$ and in addition with modified mass-energy density. The latter contained a logarithmic term, absent in the usual approach for dark matter. We studied the late time cosmological evolution of various $f(R)$ cosmological models, assuming that the universe is matter and dark energy dominated. In the usual approach for a matter and dark energy dominated universe it is assumed that matter is pressure-less, so we changed this assumption by directly putting self-interacting matter with non-zero pressure. We expressed the cosmological equations in terms of the Hubble parameter as a function of the redshift and we performed a detailed numerical analysis, which we used to study the behavior of the deceleration parameter $q(z)$ and of the effective equation of state parameter $w_{eff}$, as functions of the redshift. We compared the collisional matter $f(R)$ theories results to the ones corresponding to pressure-less matter $f(R)$ theories and to the ones coming from the $\Lambda\mathrm{CDM}$ model. After studying some very well known $f(R)$ cosmological models, as a general remark we have to note that the effect of collisional matter to $f(R)$ models is strongly model dependent. Particularly, the models appearing in equations, (\ref{oikonomod}) and (\ref{sasy}) result to $q(z)$ and $w_{eff}$ curves which are more similar to the $\Lambda\mathrm{CDM}$ curves, in comparison to the pressure-less matter $f(R)$ theories. In addition, the transition redshift that the models (\ref{oikonomod}) and (\ref{sasy}) predict are much more closer to the $\Lambda\mathrm{CDM}$ curves, again in reference to pressure-less matter $f(R)$ theories. The final picture is reversed when we consider the rest of the $f(R)$ models we studied in this article, in which case the collision-less matter $f(R)$ theory shows worst behavior than the pressure-less matter $f(R)$, in reference to the $\Lambda\mathrm{CDM}$ model. Nevertheless, the qualitative behavior of the collision-less matter $f(R)$ theory is similar to the other two models, meaning that there is a deceleration-acceleration transition. Additionally, for two of the models, we considered another form of self-interacting matter, the Cardassian matter. In this case, the model (\ref{eq:banerjee1}) actually gave results for the deceleration parameter $q(z)$ which fitted better to the $\Lambda\mathrm{CDM}$ model, in comparison to the pressure-less matter $f(R)$ model, and this is the only mentionable case corresponding to the Cardassian matter $f(R)$ theories. Furthermore, since in all the $f(R)$ models we studied there appears to be no crossing of the phantom divide, we can tentatively say that the collisional matter or Cardassian matter, cannot provide any new mechanism that can achieve a crossing of the phantom divide. Finally, we investigated in detail a very well known from the literature $f(R)$ model, that describes inflation, matter domination and late time acceleration eras. Particularly, we wanted to explicitly investigate what is the quantitative effect of collisional matter on the various evolution eras of the universe. As we demonstrated, the effect of collisional matter does not drastically affect the inflation era, does modify slightly the late time acceleration era and also enhances the matter domination era. However, the need for a fully correct description of the evolution process required the use of the modified dark energy compensate, known from the literature for it's usefulness \cite{importantpapers3}. 

It worths investigating whether there can be some overlap with theories of gravity containing non-dynamical fields in the right hand side of the Einstein equations, studied in references \cite{sot1,sot2}. These theories predict a modification of the right hand of the Einstein equations in such a way that these contain positive powers of the energy density, a feature that can be similar to the cases we studied in this article. Something like that would probably require some sort of expansion of the logarithm and of course embed the non-dynamical fields in an $f(R)$ theory context. We hope to address these issues in the future.

\section*{Acknowledgments}

V.K.O. is indebted to Ms. Saskia G. for offering I.T. and e-issues help, which significantly speeded up the completion of this article.

\end{document}